\documentclass[a4paper,14pt]{article}
\pdfoutput=1 %
\usepackage{jcappub} 
\usepackage[T1]{fontenc} 
\usepackage{color}
\usepackage{subfigure}
\usepackage{graphicx}
\usepackage{bm}

\title{\boldmath Nonlinear QED effects in X-ray emission of pulsars}

\author[a,b,d,1]{Soroush Shakeri,\note{Corresponding author.}}
\author[c]{Mansour Haghighat,}
\author[d,e]{She-Sheng Xue}

\affiliation[a]{Department of Physics, Isfahan University of Technology,\\Isfahan 84156-83111, Iran}
\affiliation[b]{ICRANet-Isfahan, Isfahan University of Technology,\\Isfahan 84156-83111, Iran}
\affiliation[c]{Department of Physics, Shiraz University,\\Shiraz 71946-84795, Iran}
\affiliation[d]{ICRANet Piazzale della Repubblica,\\10 -65122, Pescara, Italy}
\affiliation[e]{Physics Department, University of Rome La Sapienza,\\ P.le Aldo Moro 5, I00185 Rome, Italy}

\emailAdd{Soroush.Shakeri@ph.iut.ac.ir}
\emailAdd{m.haghighat@shirazu.ac.ir}
\emailAdd{xue@icra.it}

\abstract{In the presence of strong magnetic fields near pulsars, the QED vacuum becomes a birefringent medium due to nonlinear QED interactions. Here, we explore the impact of the effective photon-photon interaction on the polarization evolution of photons propagating through the magnetized QED vacuum of a pulsar. We solve the quantum Boltzmann equation within the framework of the Euler-Heisenberg Lagrangian  to find the  evolution of the Stokes parameters. We  find that linearly polarized X-ray photons propagating outward in the magnetosphere of a rotating neutron star can  acquire high values for the circular polarization parameter. Meanwhile, it is  shown that the polarization characteristics of photons besides photon energy depend strongly on  parameters of the pulsars such as  magnetic field strength, inclination angle and rotational period. Our results are clear predictions of QED vacuum polarization effects in  the near vicinity of magnetic stars which can be tested with the upcoming X-ray polarimetric observations.}

\begin{document}
\maketitle
\flushbottom
 
\section{Introduction}

Detection of vacuum nonlinear electrodynamics effects would be of profound importance for our understanding of the physical vacuum, and therefore possible tests of such effects are of great interest to a wide community of physicists. However, so far nonlinear QED effects though subject to a large number of ground experiments have not been observed. Since, the vacuum birefringence which is a manifestation of nonlinear photon-photon scattering is a very small macroscopic quantum effect \citep{1935NW.....23..246E,1936ZPhy...98..714H,1971AnPhy..67..599A,1997JPhA...30.6485H},  it is essential to have a strong enough magnetic field source in order to detect it. Although typical magnetic fields of ground magnets are relatively weak ($B\sim10^6$ G), upcoming high-intensity laser facilities  will be able to achieve ultra-high-field strengths in laboratories \citep{2012RvMP...84.1177D,2016PhyS...91b3010S,HIBEF,ELI}. At the same time the natural extreme regimes of strong magnetic fields ( of order of or above the critical QED field of $4.4 \times 10^{13}$ G) associated with pulsars and  other compact astrophysical objects, provide wide opportunities to investigate nonlinear effects of electrodynamics in vacuum.

It has been shown that QED effects have a peculiar signature on the polarization observations of pulsars, while they have a less obvious imprint on spectral parameters \citep{Ventura:1979gi,vanAdelsberg:2006uu,2011ApJ...730..131F}. It is well known that the appearance of the polarized virtual electron-positron pairs leads to different phase velocity of photons propagating in the ordinary mode (\emph{O}-mode) and the extraordinary mode (\emph{X}-mode). In fact, a pulsar is surrounded by a magnetized plasma containing relativistic electron-positron pairs (plus possibly a small amount of ions) within the light cylinder.  At low frequencies (e.g. radio waves), this plasma is birefringent giving rise to similar effects to vacuum polarization effects, but at sufficiently high frequency (e.g. X-rays) the plasma only negligibly affects the radiation as it travels through the magnetosphere \citep{1977ApJ...215L.117N,1979PhRvD..19.3565M,2003PhRvL..91g1101L,2007MNRAS.377.1095W,2006MNRAS.368.1377S,2010MNRAS.403..569W}.
It is worth mentioning that the combined plasma and vacuum polarization effects lead to a "vacuum resonance", where the contribution to the dielectric tensors of the plasma and of the vacuum compensate each other \citep{2003PhRvL..91g1101L}. In the atmospheric plasma of a strongly magnetized neutron star, vacuum polarization can induce a resonance across which a X-ray photon can convert from a low opacity mode to a high opacity mode (\emph{O}-mode or \emph{X}-mode) and vice versa, analogous to the
Mikheyev-Smirnov-Wolfenstein (MSW) mechanism for neutrino oscillation \cite{2002ApJ...566..373L,2003ApJ...588..962L}. In this mode conversion which depends on the photon energy and atmosphere density gradient, the polarization ellipse rotates $90^{\circ}$. Meanwhile, the normal modes become circularly polarized as a result of the cancellation of the plasma and vacuum effects; since both effects tend to make the mode linearly polarized in mutually orthogonal directions \citep{2003PhRvL..91g1101L,2007MNRAS.377.1095W,2009MNRAS.398..515W}. Emission from the surface of a neutron star (NS) is expected to be intrinsically polarized, because of the anisotropy introduced by the strong magnetic field \citep{1978SvA....22..214P,2006RPPh...69.2631H}. In the absence of QED vacuum polarization effects, due to  the different magnetic field orientations at the emission sites, the polarization from different regions will cancel each other leading to a drastic depolarization of the radiation collected at infinity \citep{2000ApJ...529.1011P}. It has been shown that  the vacuum polarization is also important to align the polarization vectors of the photons emitted from different patches of the NS, thus ensuring an appreciable net polarization fraction at the detector \citep{2000MNRAS.311..555H,2002PhRvD..66b3002H,2003MNRAS.342..134H}. This produces a 5-7 times larger polarization degree in the NS phase averaged signal \citep{2003MNRAS.342..134H} than in estimates where birefringence is not taken into account \citep{2000ApJ...529.1011P}.

Recently, a measurement of an optical linear polarization of the neutron star RX J1856.5-3754 has been reported  by using the European southern observatory's (ESO) Very large telescope (VLT)  that hints toward the presence of vacuum birefringence in the magnetized medium of a neutron star \citep{2017MNRAS.465..492M,Caniulef:2016vb}. RX J1856.5-3754 is brightest and youngest member of the "Magnificent Seven" (M7) which are a group of radio-quiet isolated neutron stars discovered in the soft X-rays through their purely thermal surface emission \citep{1997Natur.389..358W,2013MNRAS.429.3517M}. It has been conjectured that,  the measurement of a polarisation degree P.D. $= 16.43\%  \pm 5.26\% $ and a polarization position angle P.A. $= 145^{\circ}.39 \pm 9^{\circ}.44$ is large enough to support the presence of QED vacuum birefringence. They argued that independently on how thermal photons are produced, such a high value of linear polarization in the optical signal is extremely unlikely to be reproduced in the absence of QED vacuum polarization effects \citep{2017MNRAS.465..492M,Caniulef:2016vb}.  However, as pointed out in \citep{2006MNRAS.368.1377S} at intermediate frequencies (the optical through the infrared), the birefringence induced both by the plasma and by QED influences the observed polarization of radiation from the surface of the NS. Since the amount of plasma present varies according to the pulsar model adopted, it is hard  to make accurate predictions of plasma birefringence \citep{2003MNRAS.342..134H}. In this situation  X-ray polarimetry measurements can complement observations at optical wavelengths and are very significant to  discriminate between the case of a gaseous atmosphere and a condensed  surface emission models  \citep{2017MNRAS.465..492M,Caniulef:2016vb}. As discussed in \citep{2011ApJ...730..131F}, vacuum polarization dominates up to distance of about 3000 times the stellar radius for keV photons under the typical conditions of a magnetar magnetosphere, outside this radius plasma polarization starts to dominate. 

In spite of having a vast amount of polarization data in the optical and radio bands, the only successful polarization measurement in X-rays dates back to 1976, when a Bragg polarimeter onboard OSO-8 measured the polarization of the Crab nebula which is one of the brightest object in the X-ray sky \cite{1972ApJ...174L...1N,1978ApJ...220L.117W}.
Polarimetry based on the classical techniques, Bragg diffraction and Compton
scattering (which require rotation of instrument), became seriously mismatched with an enormous increase of the sensitivity in imaging and spectroscopy devices \citep{2010NIMPA.623..766B,2010xpnw.book.....B}. As a result, no polarimeters were included in major X-ray missions
by the NASA or ESA (e.g. Einstein, Chandra and XMM-Newton). The advent of a new generation of detectors, with the development of sensors based on the photoelectric effect \citep{2001Natur.411..662C}, has renewed interest in X-ray polarimetry leading to the several polarimetric missions recently proposed. The X-ray Imaging Polarimetry Explorer (XIPE) \citep{2013ExA....36..523S}, the Imaging X-ray Polarimetry Explorer (IXPE) \citep{2013SPIE.8777E..0WA}, and the Polarimeter for Relativistic Astrophysical X-ray Sources (PRAXyS) \citep{2015AAS...22521401W} will allow  to exploit strongly magnetized NS as a laboratory for fundamental physics and eventually probe the properties of the matter in the strong-field regimes.  

Generally, photon interaction with a charged particle causes the outgoing photons to be linearly polarized with  non-vanishing values for \emph{Q} and/or \emph{U}, but there is no way to generate circular polarization parameter \emph{V} in a scattering process of linear Maxwellian electrodynamics. Significant circular polarization can be generated through the conversion of linear polarization of radiation by taking into account several interactions, such as Compton scattering in a background  whether the external magnetic field or noncommutative space time \citep{2010PhRvD..81h4035Z,2017JHEP...02..003T}, photon-neutrino interaction \citep{Mohammadi:2013ksa,Khodagholizadeh:2014nfa}, and photon-photon scattering \citep{2014PhRvA..89f2111M,Motie:2011az,2017PhRvA..95a2108S}. Here we focus on the nonlinear aspects of QED in the perturbative regime that induces an effective interaction between photons. Recently, we have  presented a formalism in  \cite{2017PhRvA..95a2108S} to consider the polarization effects caused by the photon-photon interaction in the presence of  background electromagnetic field. Based on this formalism, we are going to investigate the effects of nonlinear QED interactions in the presence of strong magnetic field associated with pulsars. Here, we focus on the soft X-ray photons propagating through the magnetosphere of a pulsar.
In the following we neglect the plasma contribution to the polarization properties which is  a reasonable assumption in the high-frequency range. The aim of this paper is to explore all the  relevant parameters  and identify specific signatures in the polarization of outcoming photons from the pulsar. 

The paper is organized as follows: Sec. \ref{NM} begins with introducing the set of equations that describe the time evolution of the Stokes parameters in the presence of a magnetic field, taking into account the Euler-Heisenberg effective Lagrangian. We then  solve numerically these equations in order to find polarization characteristics of the original linearly polarized X-ray photons, which are propagating outward through decreasing magnetic field of a pulsar. In Sec. \ref{CAP} we consider the generation of the circular polarized X-ray emission from the polar cap and possible extension to more general situations. Moreover, we discuss the implication of our results for the X-ray polarization signals from pulsars in more realistic configurations. Finally, in the last section we give some concluding remarks.

\section{Pulsar Polarization due to Nonlinear QED effects}
\label{NM}
\begin{figure}[tbp]
\centering 
\includegraphics[width=3in]{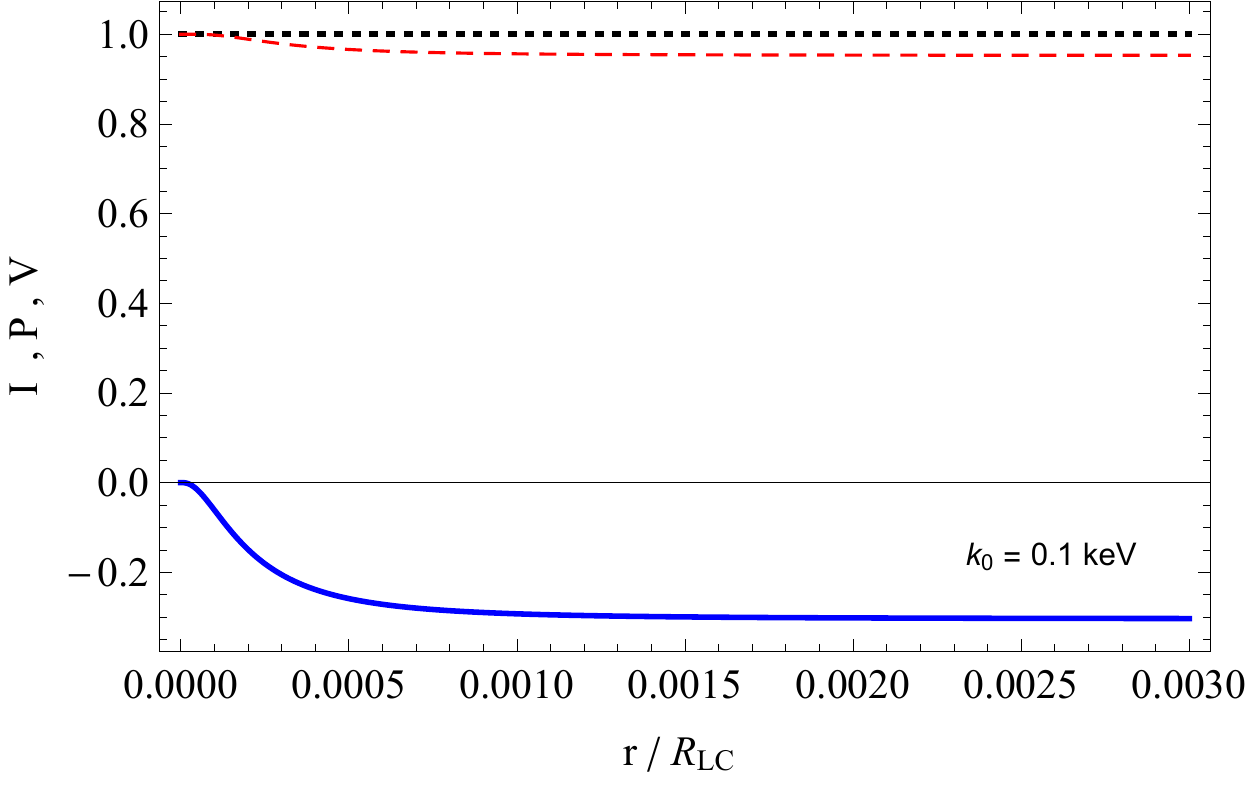}
\hfill
 \includegraphics[width=3in]{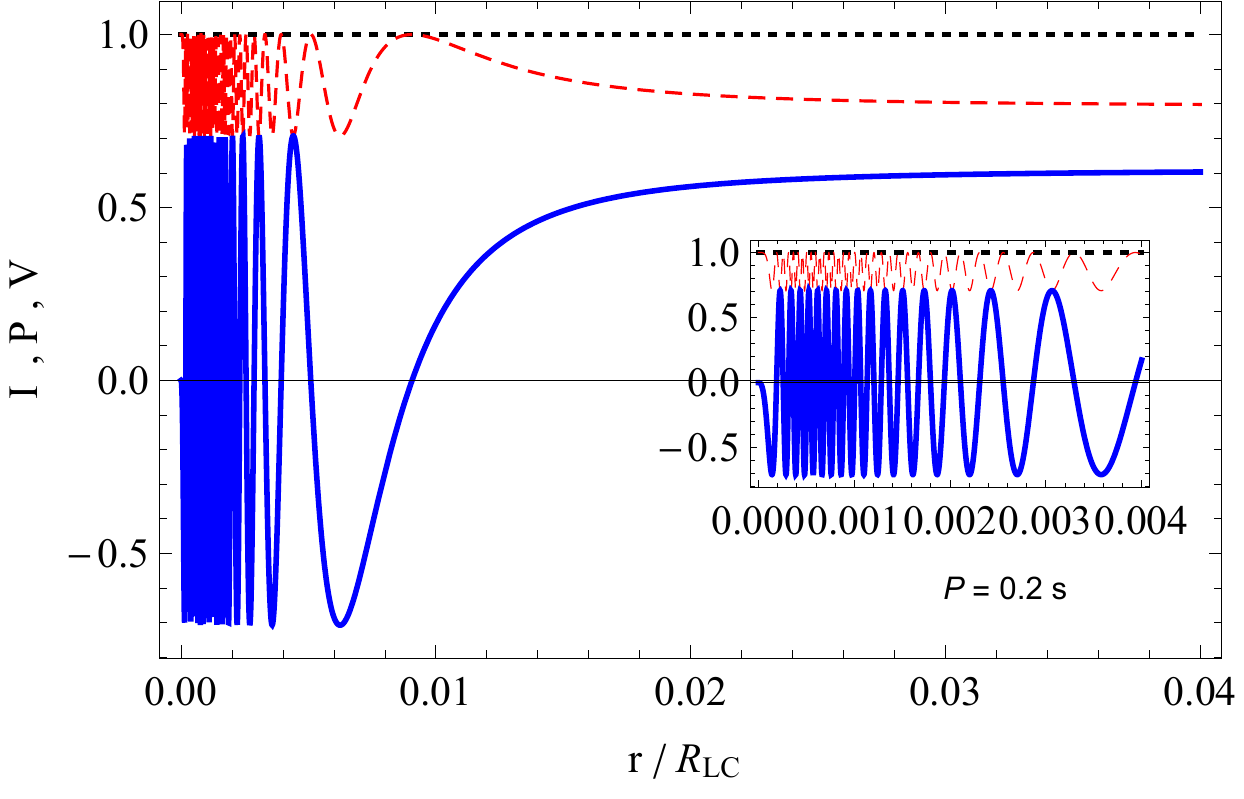}
 \hfill
\includegraphics[width=3in]{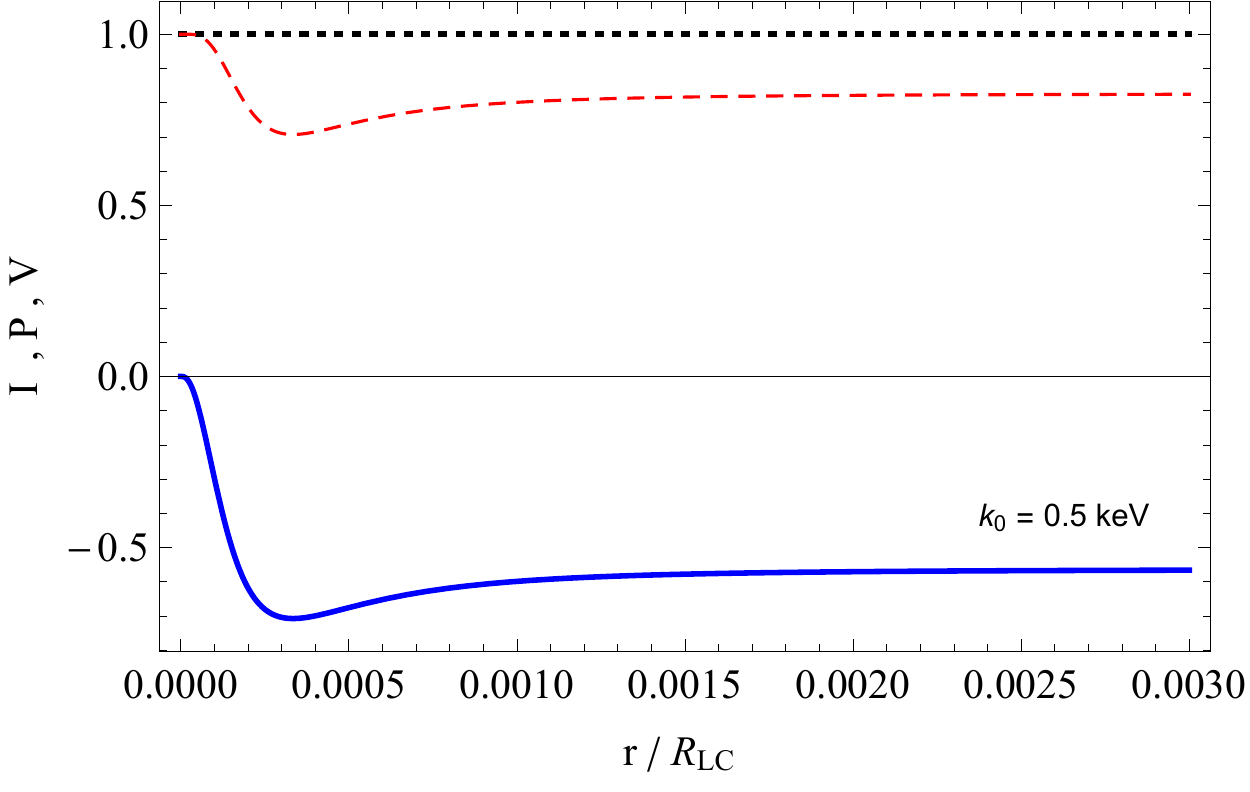}
\hfill
\includegraphics[width=3in]{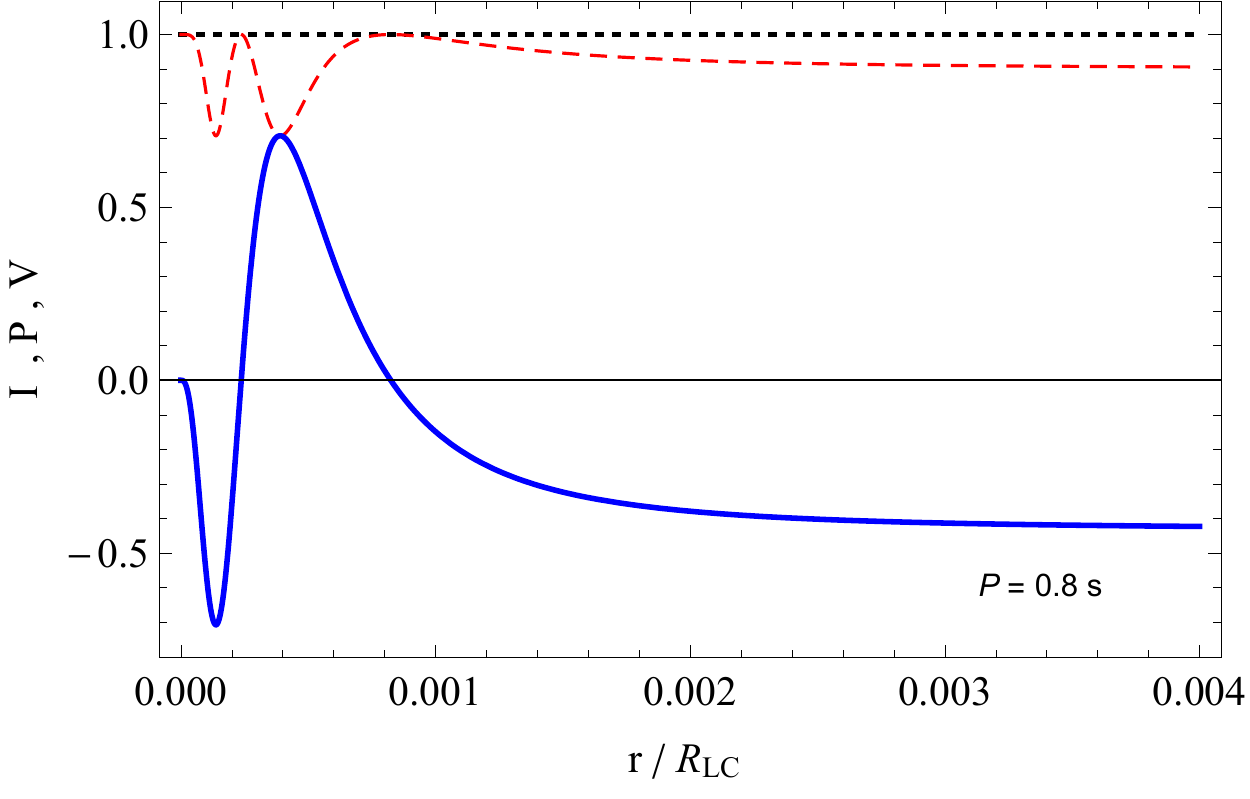}
\hfill
\includegraphics[width=3in]{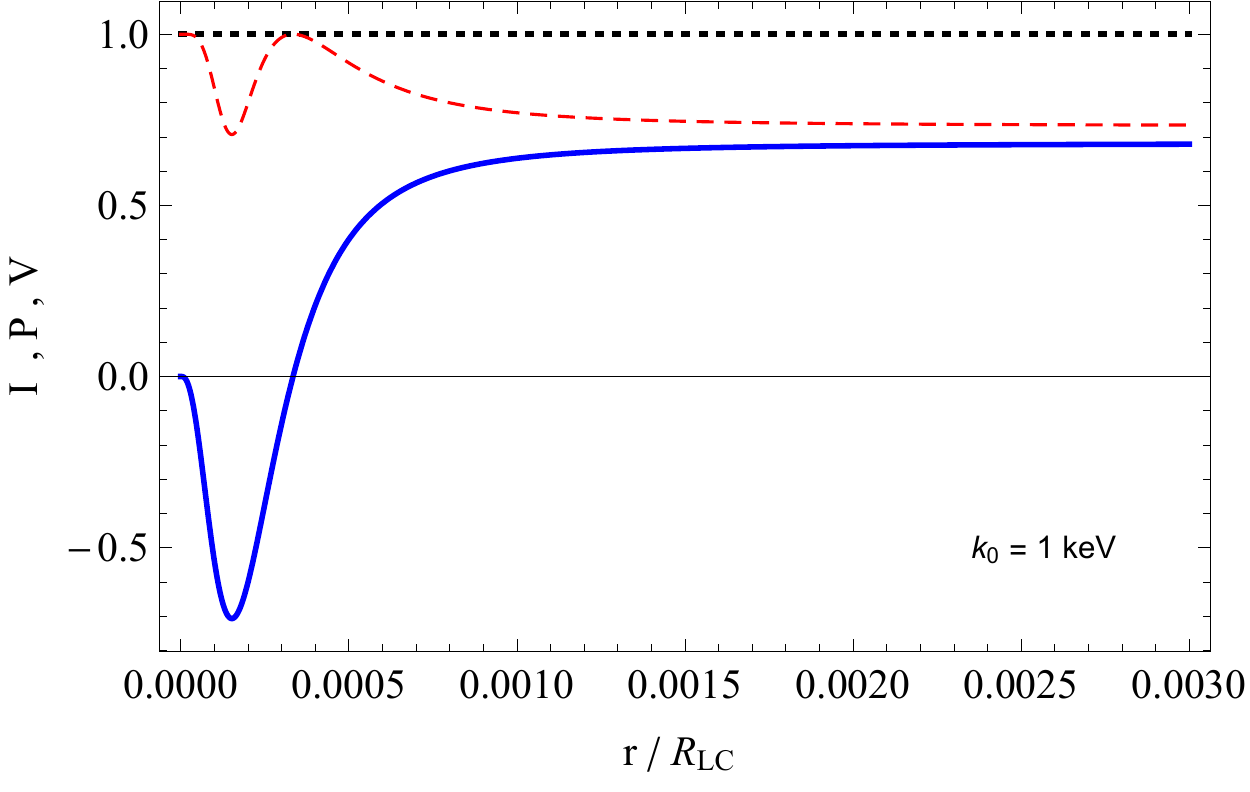}
 \hfill
\includegraphics[width=3in]{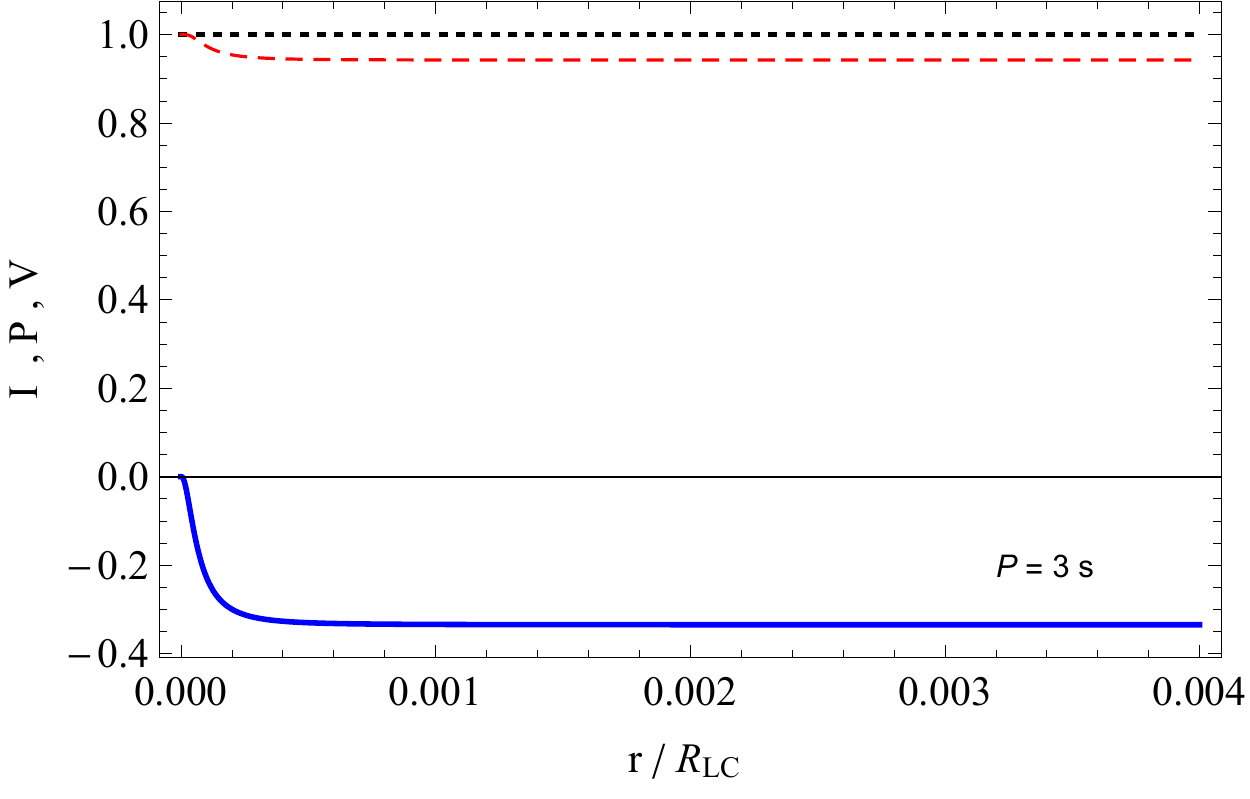}

    \caption{\label{f01} Evolution of the dimensionless  Stokes parameters I [dotted (black)], P [dashed (red)] and V [solid (blue)] for a particular photon moving radially
outward in the magnetosphere. The horizontal axis  $r/R_{LC}$ is the photon distance away  from the stellar surface (emission point) in units of the light-cylinder radius.  In the left panels, the different diagrams are plotted for various photon energies $k_{0}$ as labeled and rotational period $P= 1s$. In the right panels, they are plotted   for various rotational period $P$  as labeled and  photon energy $k_{0}= 1 keV$. In this cases, the other parameters are surface magnetic field $B_{0}=2\times10^{12} G$ and inclination angle $\alpha= 15^{\circ} $.}
     \end{figure}

\begin{figure}[tbp]
\centering 
\includegraphics[width=3in]{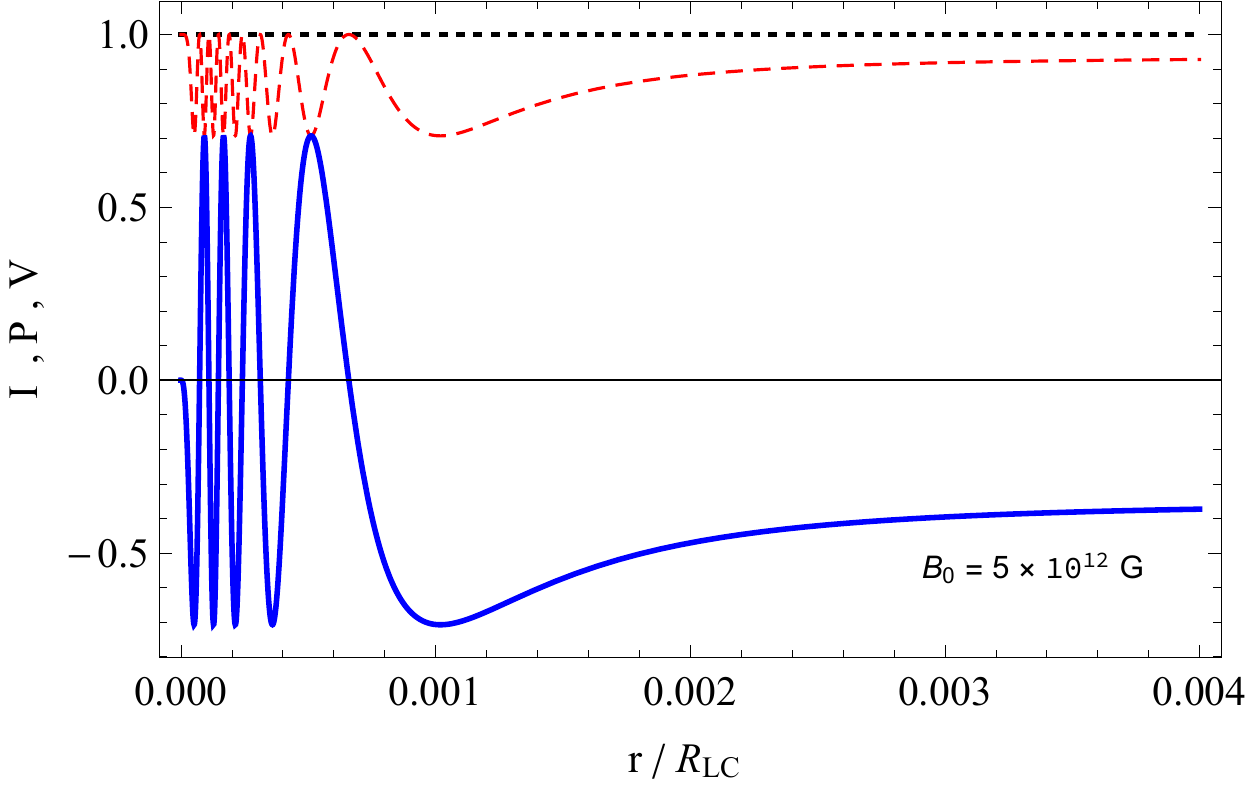}
\hfill
 \includegraphics[width=3in]{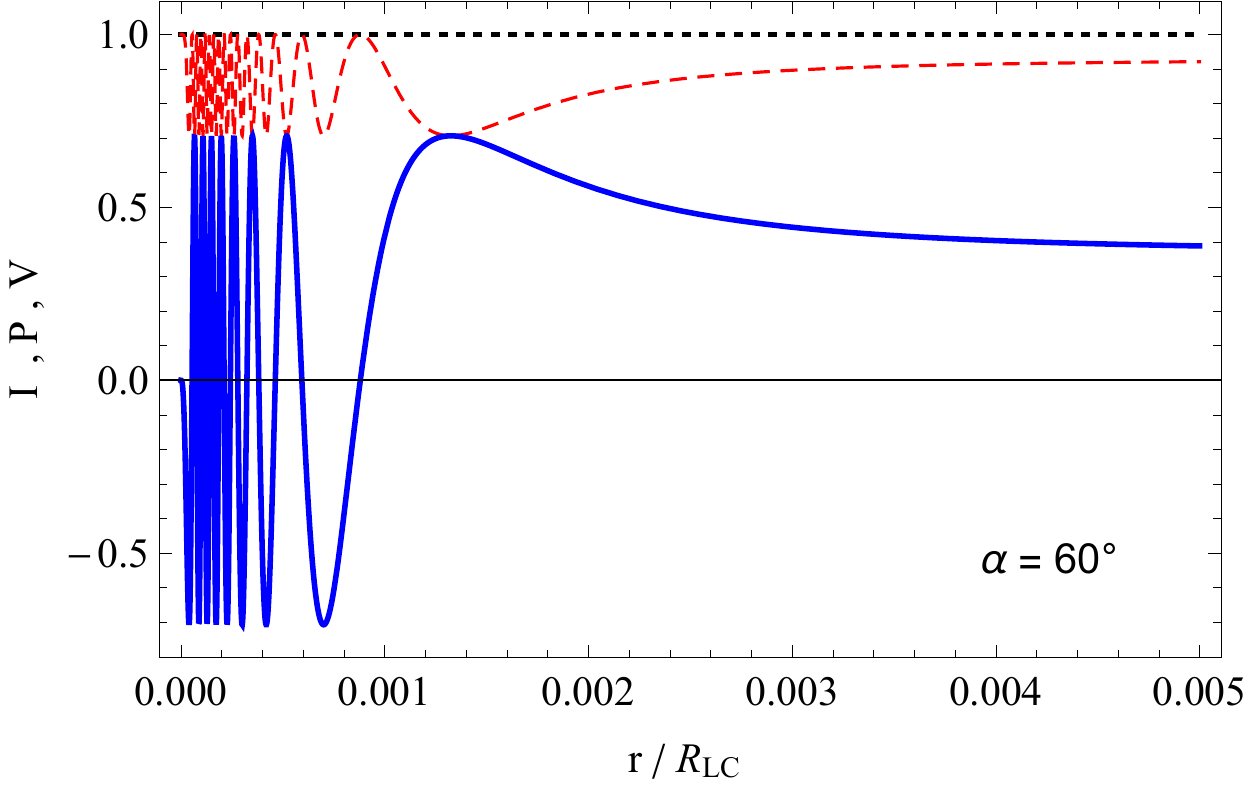}
 \hfill
\includegraphics[width=3in]{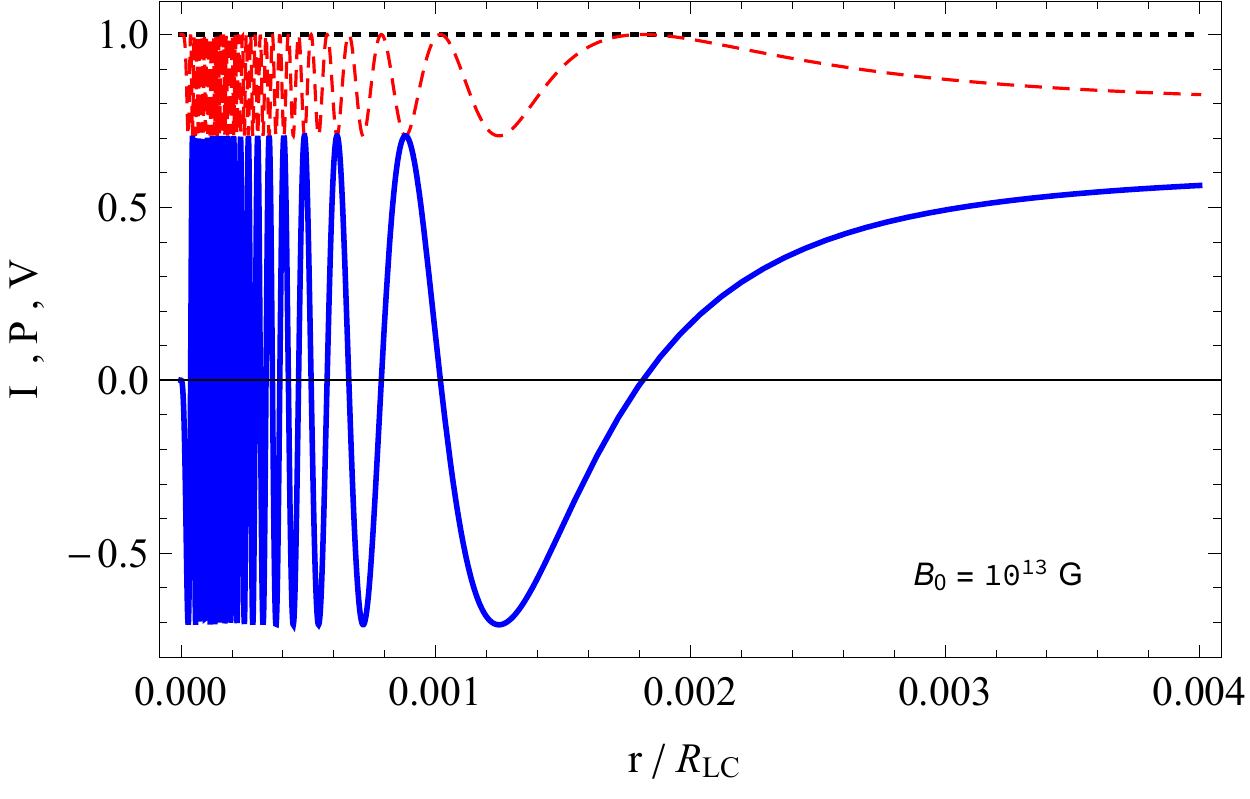}
\hfill
\includegraphics[width=3in]{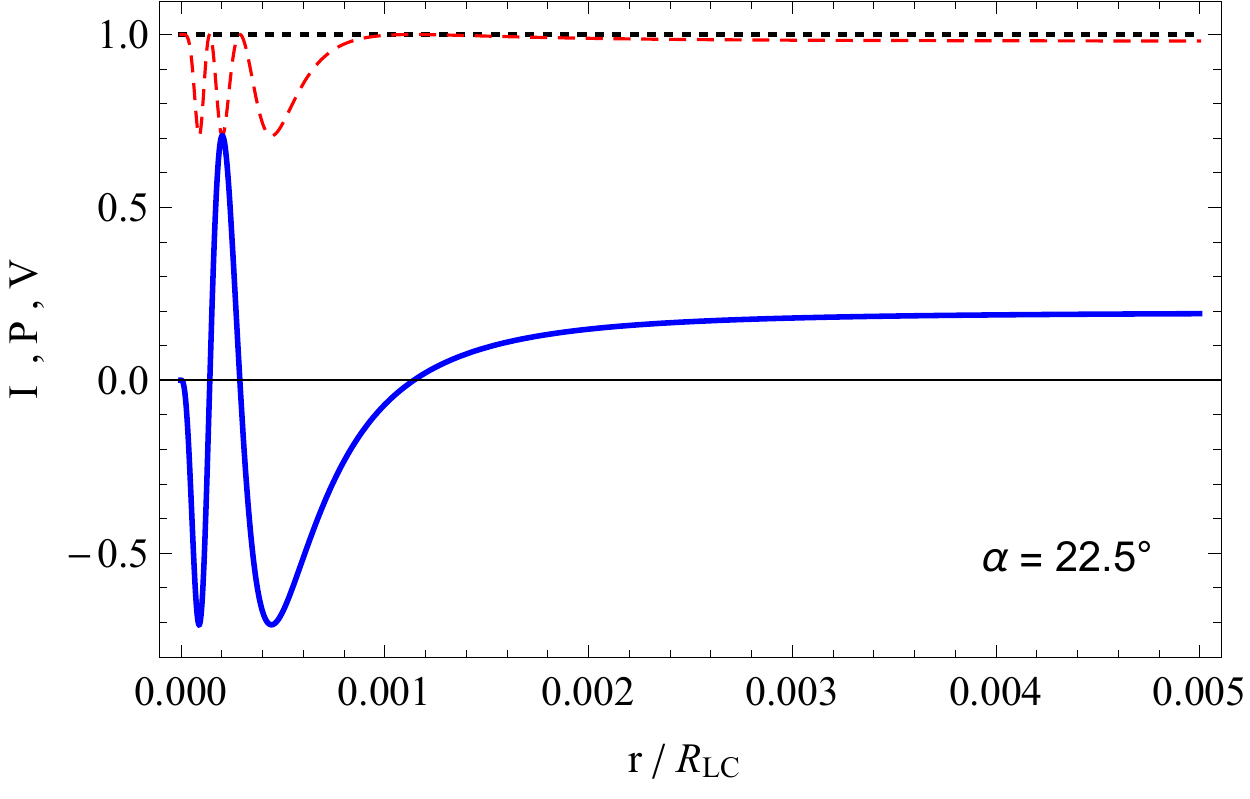}
\hfill
\includegraphics[width=3in]{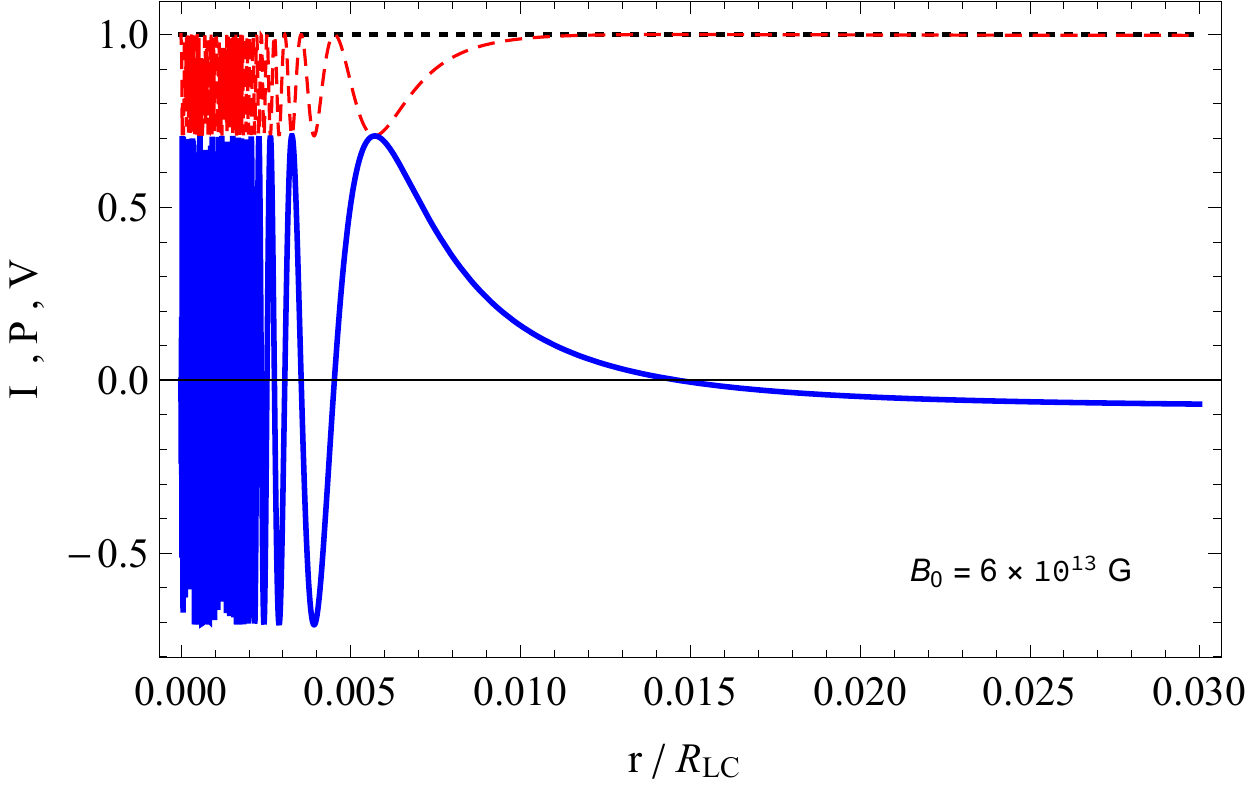}
 \hfill
\includegraphics[width=3in]{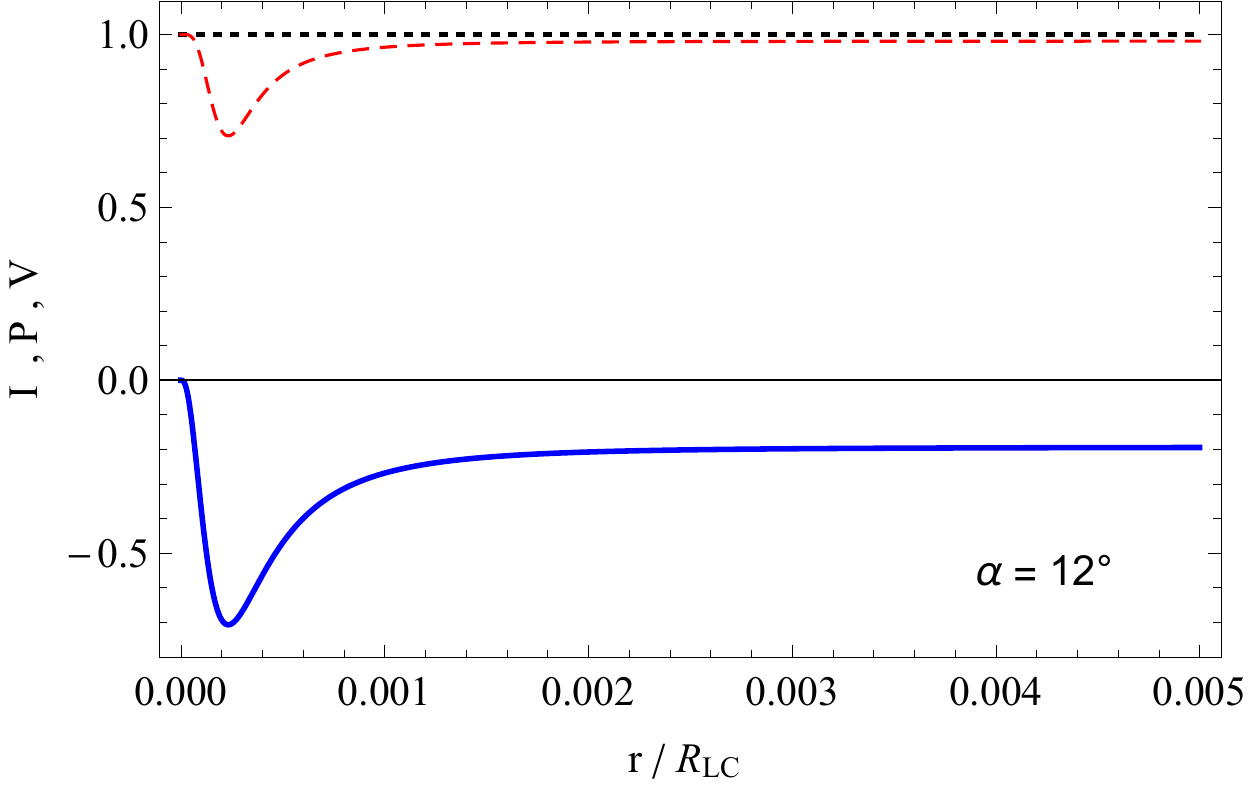}

    \caption{\label{f04} Same as Fig. \ref{f01}, except for different surface magnetic field $B_{0}$ as labeled and the same inclination angle $\alpha= 15^{\circ}$ (left panels), and  different  inclination angle $\alpha$ as labeled and the same surface magnetic field $B_{0}=2\times10^{12} G$ (right panels). Both are plotted for photon energy $k_{0}= 1 keV$ and rotational period $P = 1 s$.}
     \end{figure}

A convenient way to describe the polarization characteristics of a nearly monochromatic electromagnetic radiation emitted by a source is through the four Stokes parameters \emph{I}, \emph{Q}, \emph{U} and \emph{V}   \citep{Chandrasekhar:1960tz,Jackson:1975up,Rybicki:2008vo}. The parameter \emph{I} gives the intensity of the radiation,  the parameters \emph{Q} and \emph{U}  describe the linear polarization, and  the  parameter \emph{V} is a measure of the circular polarization. Meanwhile, the total degree of linear polarization can be represented by $P_{L}=\sqrt{Q^{2}+U^{2}}$. For a general source of light, the four stokes parameters satisfy the general relation $I^{2}\geq Q^{2}+U^2+V^2$, where the equality holds for $100 \%$ polarized radiation.  Our approach is based on a quantum-mechanical description of the Stokes parameters and their time evolution  are given by so-called quantum Boltzmann equation \citep{1996AnPhy.246...49K,2009PhRvD..79f3524A}. In the following,  in order to describe the nonlinear dynamics of electromagnetic fields in vacuum, we adopt the Euler-Heisenberg effective Lagrangian taking into account one-loop QED quantum corrections \citep{1935NW.....23..246E,1936ZPhy...98..714H,1936AnP...418..398E,1998PhRvD..57.2443D,Dunne:2012hp}. The time evolution of the the Stokes parameters due to the Euler-Heisenberg  effective Lagrangian, has been extensively discussed  recently in Ref. \citep{2017PhRvA..95a2108S} in the presence of slowly varying background electromagnetic fields (electric and magnetic fields). In the present paper, we apply the same formalism in the case of a pure background magnetic field, where magnetic-field components are the only non-zero components of the background field strength tensor [See Eqs. (47)-(50) of Ref. \citep{2017PhRvA..95a2108S}]. Consequently the time evolution for the Stokes parameters  can be obtained 
\begin{align}\label{e38}
\dot I=\dot \rho_{11}+\dot \rho_{22}=0,
\end{align}
\begin{align}\label{e39}
\dot Q=\dot \rho_{11}-\dot \rho_{22}=-\frac{4\alpha^{2}k_{0}}{15m^{4}_{e}} [\vec B\cdot(\hat \epsilon_{2}\times\hat k)\ \vec B\cdot(\hat \epsilon_{1}\times\hat k )]V,
\end{align}
\begin{align}\label{e40}
\dot U=\dot \rho_{21}+\dot \rho_{12}=-\frac{2\alpha^{2}k_{0}}{15m^{4}_{e}} [(\vec B\cdot\hat \epsilon_{2}\times\hat k)^{2}-(\vec B\cdot\hat \epsilon_{1}\times\hat k)^{2}]V,
\end{align}
\begin{align}\label{e41}
\dot V=i(\dot \rho_{12}-\dot \rho_{21})=\frac{2\alpha^{2}k_{0}}{15m^{4}_{e}} [((\vec B\cdot\hat \epsilon_{2}\times\hat k)^{2}-(\vec B\cdot\hat \epsilon_{1}\times\hat k)^{2})U+2(\vec B\cdot\hat \epsilon_{2}\times\hat k\ \vec B\cdot\hat \epsilon_{1}\times\hat k\ )Q].
\end{align}
We choose $\hat k$ in the direction of outcoming photons from the pulsar with real polarization vectors in the transverse condition ($\hat \epsilon_{1}(k)$ and  $\hat \epsilon_{2}(k)$ $\perp$ $\hat k$). From Eq. (\ref{e38}) it is clear that the total intensity of photons does not depend on the photon-photon forward-scattering term, since there is no dissipation in QED below the threshold energy of pair production. According to Eqs. (\ref{e39})-(\ref{e41}), as a result of nonlinear QED interactions, the linear polarization of a radiation (\emph{Q} and/or \emph{U} $\neq$ 0) can be converted into the circular polarization (\emph{V} $\neq$ 0) proportional to ($\alpha^{2} B^{2} k_{0} /m^4_{e}$) due to the propagation in the presence of  magnetic field. Both the energy dependence of the polarization features  presented in this approach, and the geometry of the magnetic field with respect to the outgoing photons affect the time evolution of photons in the photon ensemble. In order to solve the coupled Eqs. (\ref{e38})-(\ref{e41}) in the case of a particular photon across the magnetosphere of a pulsar, we rewrite them as follows:

\begin{align}\label{e42}
\dot I=0, \ \ \  \  \dot Q=-\Omega_{QV}V, \ \ \ \  \dot U=-\Omega_{UV}V, \ \  \  \ \dot V=\Omega_{QV}Q+\Omega_{UV}U,
\end{align}
where
\begin{align}\label{e43}
\Omega_{QV}=\frac{4\alpha^{2}k_{0}}{15m^{4}_{e}} [\vec B\cdot(\hat \epsilon_{2}\times\hat k)\ \vec B\cdot(\hat \epsilon_{1}\times\hat k )], \ \ \  \Omega_{UV}= \frac{2\alpha^{2}k_{0}}{15m^{4}_{e}} [(\vec B\cdot\hat \epsilon_{2}\times\hat k)^{2}-(\vec B\cdot\hat \epsilon_{1}\times\hat k)^{2}].
\end{align}
According to Eq. (\ref{e42}), $\ddot V$ can be cast into
\begin{align}\label{e44}
\ddot V=-\Omega ^{2}V+\dot \Omega _{QV} Q+\dot \Omega _{UV} U \ \  \  \ \ \ \ ; \ \ \ \ \ \ \Omega ^{2}=\Omega^{2} _{QV} +\Omega^{2} _{UV}. 
\end{align}
 Now in order to find the Stokes parameters, we have to determine the several free parameters in Eqs. (\ref{e42}) and (\ref{e43}), such as the magnetic field vector  $\vec B$, the polarization vectors ($\hat \epsilon_{1},\hat \epsilon_{2}$) and the energy of the outgoing photons ($k_{0}=h\nu_{0}$). In this respect  we assume the magnetic field of pulsar is dominated by a dipole field structure and pulses are emitted from the open field lines associated with the magnetic poles. Besides, the magnetic axis of a pulsar is  not necessary the same as its rotational axis.  The angle between the magnetic dipole axis and the spin axis is called the inclination angle $\alpha$ which is one of the free parameters with a range $\alpha\sim10^{\circ}-30^{\circ}$ for  the majority of pulsars.  We also note that the exact determination of the polarization vectors is not possible, only $ \hat k\cdot\hat \epsilon=0$ fixes the directions of the two polarization components orthogonal to the direction of the emitted photon. Moreover, it has been supposed that initially when a photon is emitted, $\hat k$ and $\vec B$ have the same directions and through the pulsar rotation, the magnetic field axis $(\vec B$) rotates around the rotational axis with a relative angle $\alpha$. 
Therefore we define;
\begin{align}\label{e45}
\vec B=B(r(t))\begin{pmatrix} \sin\alpha\cos(\omega t) \\
\sin\alpha\sin(\omega t) \\
\cos \alpha \\
\end{pmatrix}, \ \  \hat k=\begin{pmatrix} \sin\alpha \\
 0 \\
\cos \alpha \\
\end{pmatrix},\  \ \hat \epsilon_{1}^{}=\begin{pmatrix} \cos\alpha \\
 0 \\
-\sin \alpha \\
\end{pmatrix}, \ \ \hat \epsilon_{2}^{}=\begin{pmatrix} 0 \\
 1 \\
0 \\
\end{pmatrix},
\end{align}
where $\omega=2\pi/P$ is the pulsar's angular velocity and the polarization vectors are chosen  to be orthogonal to the emission direction. The magnetic field has magnitude $B(r(t))=B_{0}(R_{0}/r(t))^{3}$, where $B_{0}$ is the magnitude of the (dipole) surface field at the magnetic pole,  $R_{0}=10\ \mathrm{km}$ is the typical radius of the pulsar, and $r(t)$ is the distance from the center of the star. From set of parameters Eq. (\ref{e45}) for $\Omega _{QV} $ and $\Omega _{UV} $ we  obtain:
 \begin{align}\label{e46}
 \Omega_{QV}=2G[\sin^{2}\alpha\cos\alpha\ \sin\omega t-\sin^{2}\alpha\cos\alpha\sin\omega t\cos\omega t],
 \end{align}
 and
 \begin{align}\label{e47}
 \Omega_{UV}=G[(\sin\alpha\sin\omega t)^{2}-(\sin\alpha\cos\alpha)^{2}\ (\cos\omega t-1)^{2}],
 \end{align}
 where
 \begin{align}\label{e48}
 G=\alpha \frac{k_{0}}{30 \pi}[\frac{B(r)}{B_{c}}]^{2}=3.4\times10^{10}[\frac{k_{0}}{1keV}][\frac{B(r)}{10^{10}G}]^{2}s^{-1}.
 \end{align}
 in which $B(r)$ and  $k_{0}$ are given in Gauss and keV, respectively, and also $B_{c}$=$m_{e}^{2}/e$=$1.282\times10^{13}$ G is the critical magnetic field. It has been assumed that  the initial value of the  circular polarization $V_{0}=0$ and completely linearly polarized radiation  $P_{0}=\sqrt{Q_{0}^{2}+U_{0}^{2}}=1$ ($Q_{0}=U_{0}={1}/{\sqrt{2}}$), where the Stokes parameters are normalized by the intensity \emph{I}, giving raise to dimensionless quantities. Regarding this set of parameters, we use a numerical integration to find the evolution of the Stokes parameters along the radial ray from the emission point on the NS surface up to large distances where the values of the Stokes parameters are ‘frozen’ and no longer evolve. Here the effects of gravitational light bending are neglected, however it is straightforward to incorporate this effect into the model \citep{vanAdelsberg:2006uu,2002ApJ...566L..85B}. The  results for different parameter values such as the energy of photons ($k_{0}$), the rotational period of the star (\emph{P}), the surface magnetic fields ($B_{0}$) and the inclination angles ($\alpha$) are displayed in Figs. \ref{f01} - \ref{f04}. These figures show the evolution of the Stokes parameters relates to a particular ray as a function of relative distance $r(t)/R_{LC}$, where $R_{LC}=c/\omega$ is the radius of the light cylinder.

As it is shown, the conversion rate of  linear polarization of radiation to circular polarization depends strongly on the  coupling strength parameters $\alpha^{2} B^{2} k_{0} /m^4_{e}$ (See Eqs. (\ref{e42}) and (\ref{e43})). The  conversion becomes more significant as the coupling strength is increased which manifest as rapid oscillations in Stokes parameters. As the photon moves away from the star surface, the magnetic field orientation rotates and its value decreases, hence the photon polarization is affected by the magnetic field as a function of space and time. Consequently the direction of the electric field of each photon varies inside this region. The polarization direction is eventually frozen, as photons travel  far enough from the surface of the star inside the light cylinder, where the magnetic field is relatively weak with respect to the surface field $B_{0}$.

Since the coupling is weaker at lower frequencies, Fig. \ref{f01} (left panels) pointed out the less energetic photons obtain smaller values of circular polarization parameter V during their propagation. Here we have considered soft X-ray emission in which  the isolated neutron stars (e.g. M7) are bright enough for the future X-ray polarimeters operating in this energy range \citep{2014mbhe.conf..288M} to test QED effects in the strong-field regime of NSs. As one can find from Fig. \ref{f01} (right panels)  the rotational-period variation of the star does change the polarization properties. Our results, in agreement with those of Refs. \citep{vanAdelsberg:2006uu,2000MNRAS.311..555H} show that substantial circular polarization can be generated via a sufficiently rapid rotation. While previous considerations have relied on solving the transfer equation of the polarization modes or  Stokes parameters when the approximations
of geometrical optics are used and viewing fields as classical fields \citep{Kubo:1983fk,vanAdelsberg:2006uu,2000MNRAS.311..555H}, our quantum mechanic treatment is based on quantum Boltzmann equation to evaluate Stokes parameters.

Unlike our case, in the pervious  studies  \citep{vanAdelsberg:2006uu,Taverna:2015uu,2003MNRAS.342..134H} $U_{0}=0$, $V_{0}=0$ and $Q_{0}=1$ have been applied as the initial values to evaluate the Stokes parameters. In their considerations the polarization of the radiation remains in one of the two polarization modes ( e.g. $Q_{0}=1$ $(Q_{0}=-1)$ marked as \emph{O}-mode (\emph{X}-mode) ) up to the polarization-limiting radius, where the polarization direction is frozen, and a circular polarization degree can arise only as a consequence of the polarization mode evolution at this radius. But in this respect we show that circular polarization V can be developed through the conversion of linear polarization of radiation even far below polarization-limiting radius and photons can obtain high values of V during  their propagation through the magnetic field. Besides,  Fig. \ref{f01} (right panels)  shows that for smaller values of  rotational period (top panel), the polarization-limiting radius is a large fraction of the radius of the light cylinder, the same is true for the higher values of the surface magnetic field [See Fig. \ref{f04} (left panels)]. We produce similar plot by varying the inclination angle which is presented in Fig. \ref{f04} (right panels).

\begin{figure}[tbp]
\centering 
\includegraphics[width=3in]{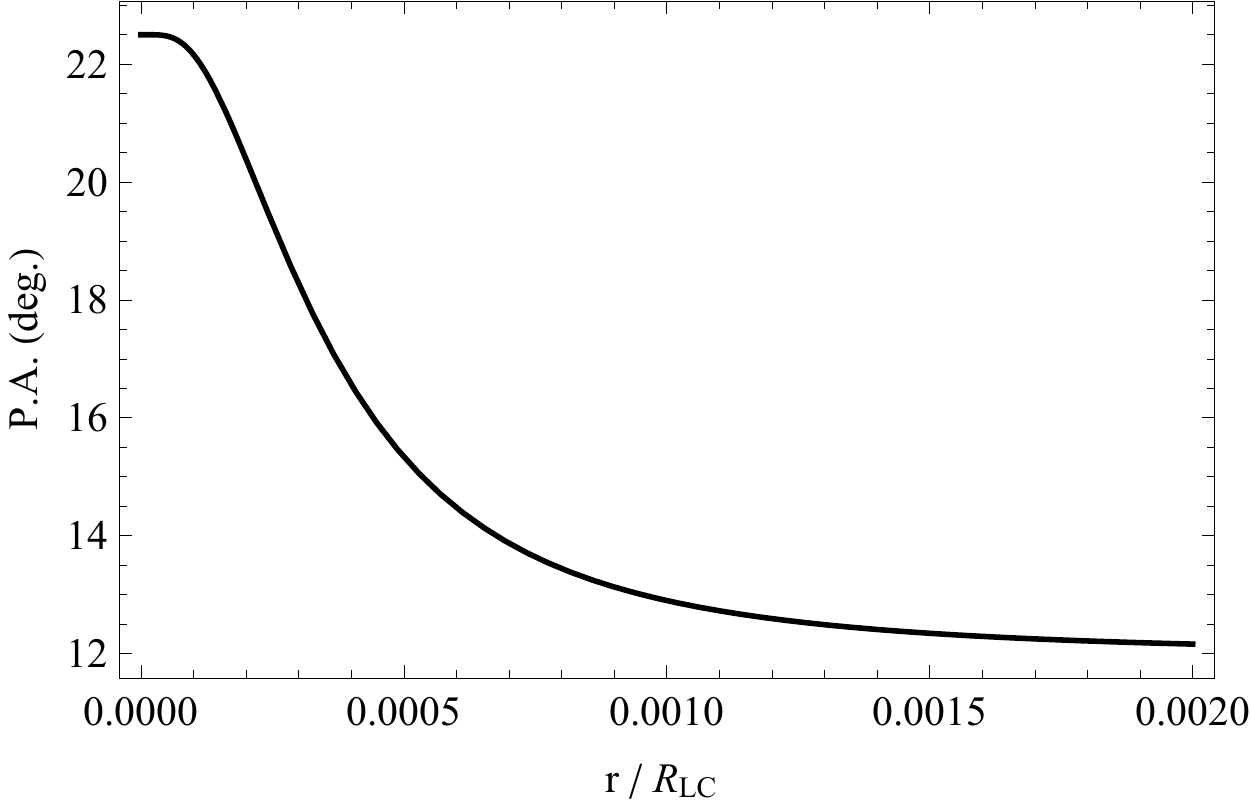}
\hfill
 \includegraphics[width=3in]{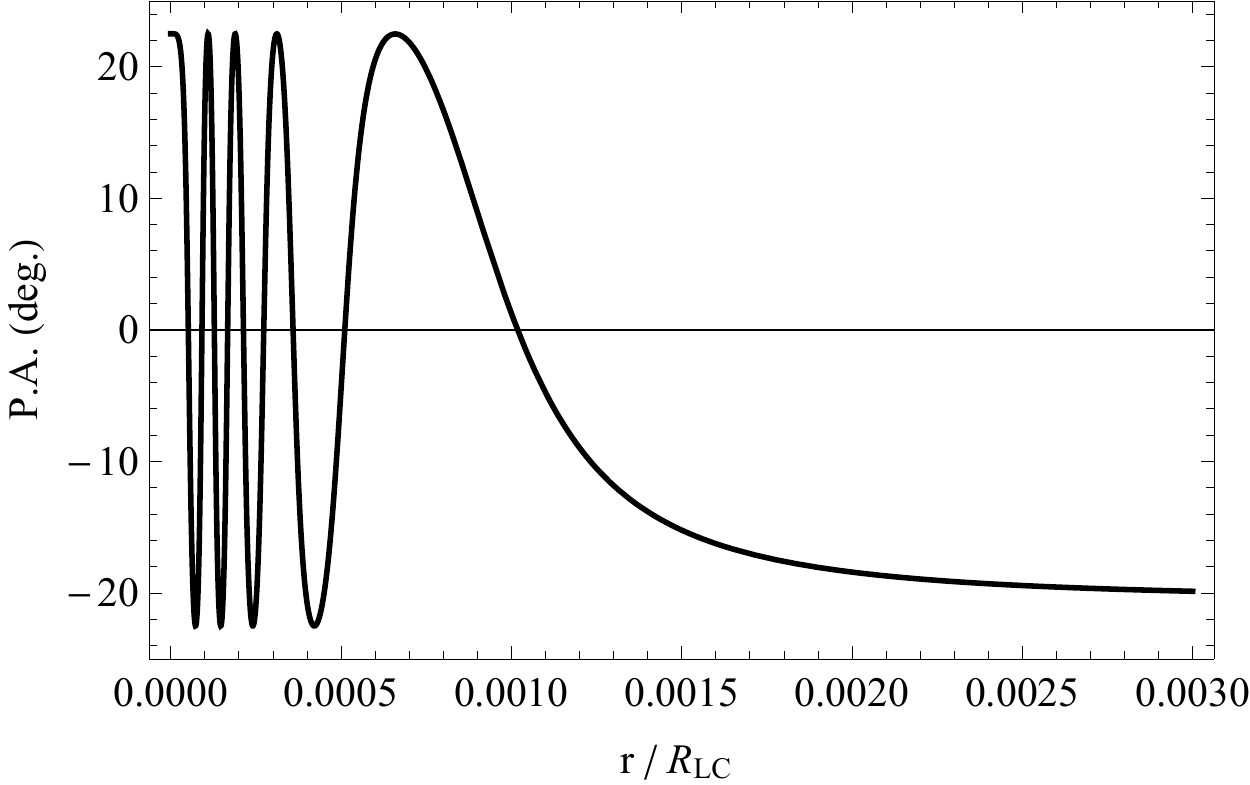}
 \hfill
\includegraphics[width=3in]{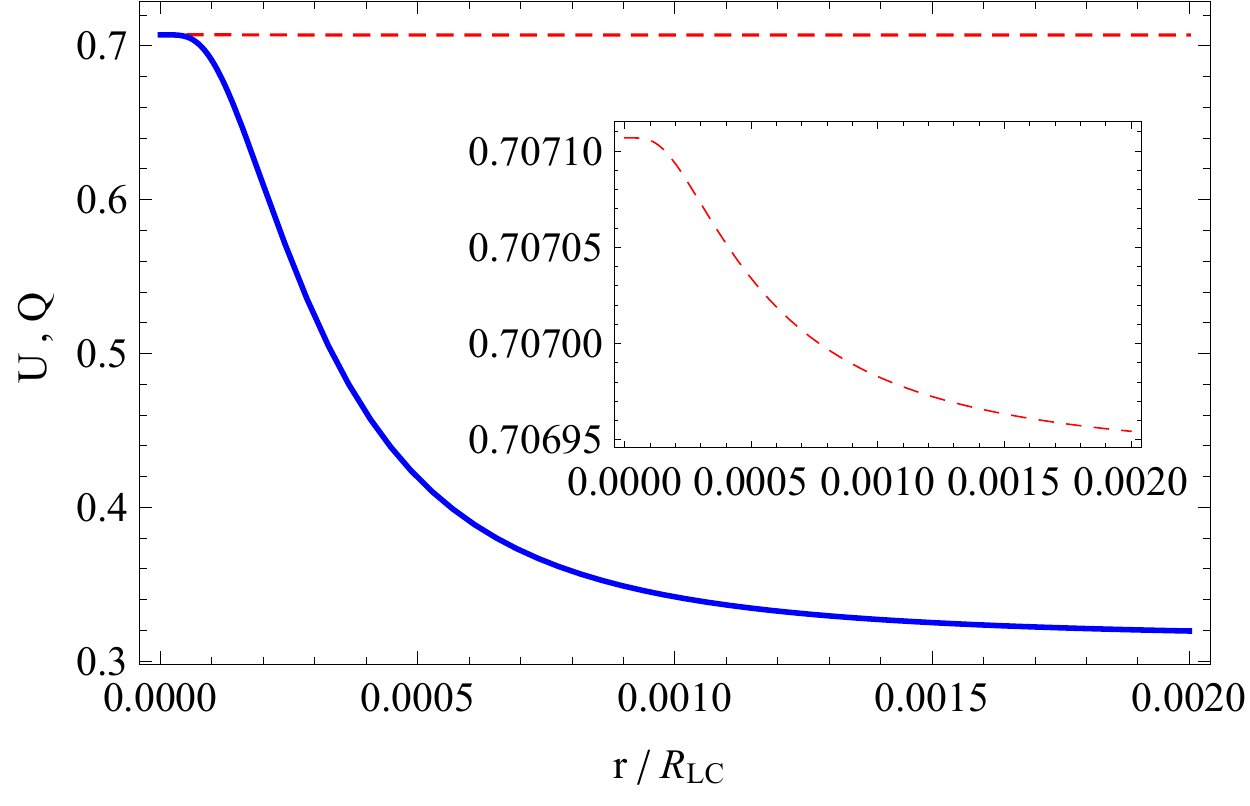}
\hfill
\includegraphics[width=3in]{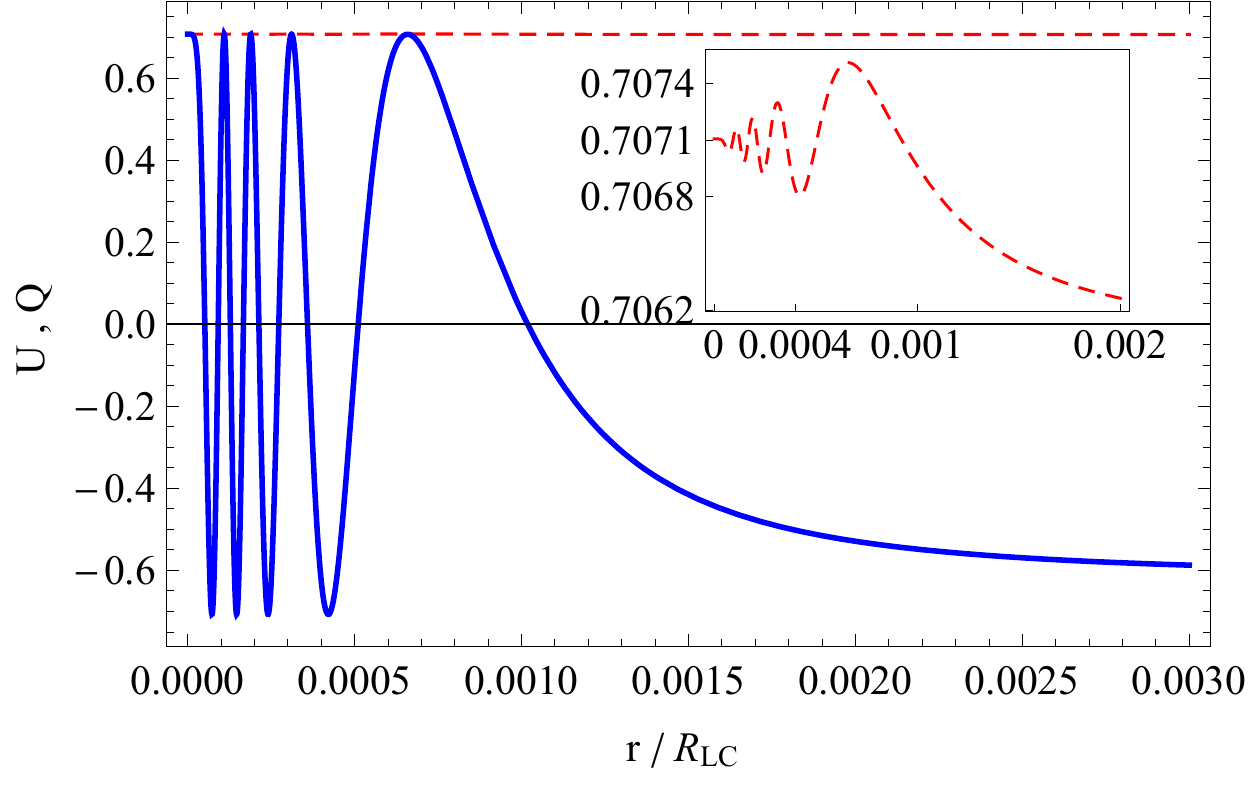}
\hfill
\includegraphics[width=3in]{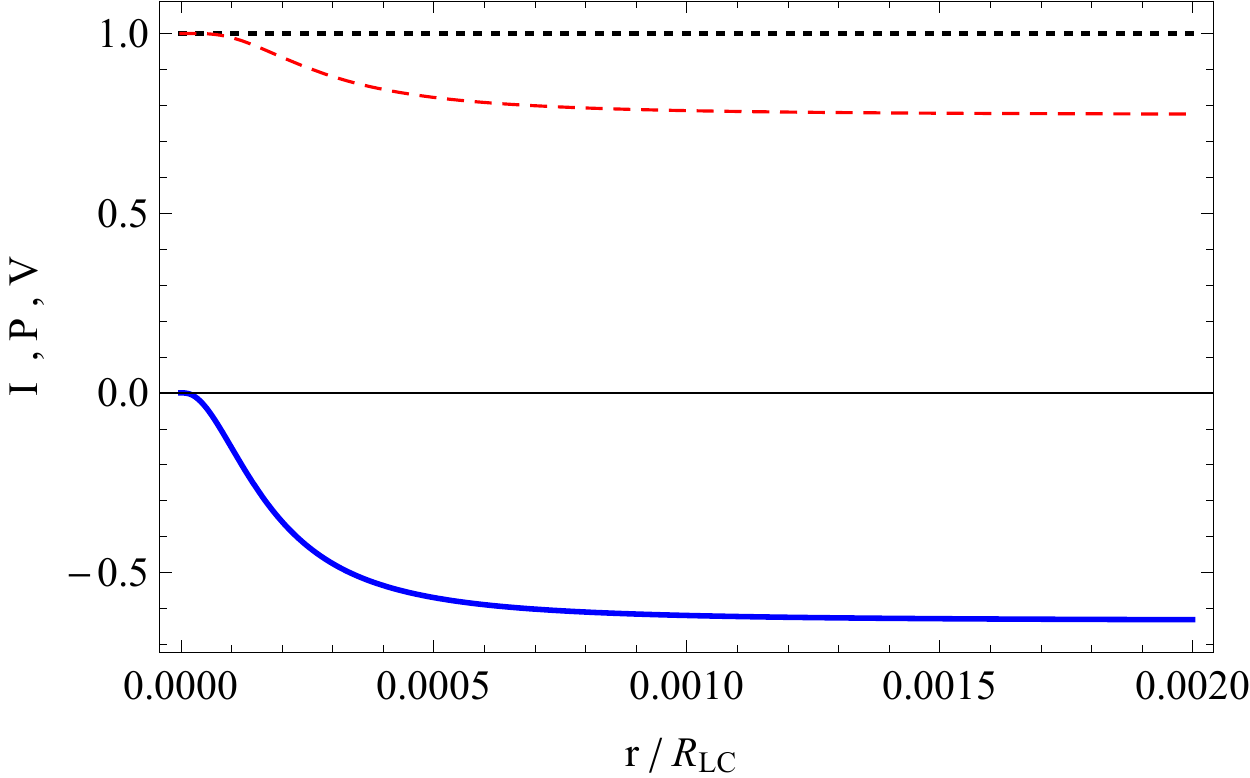}
 \hfill
\includegraphics[width=3in]{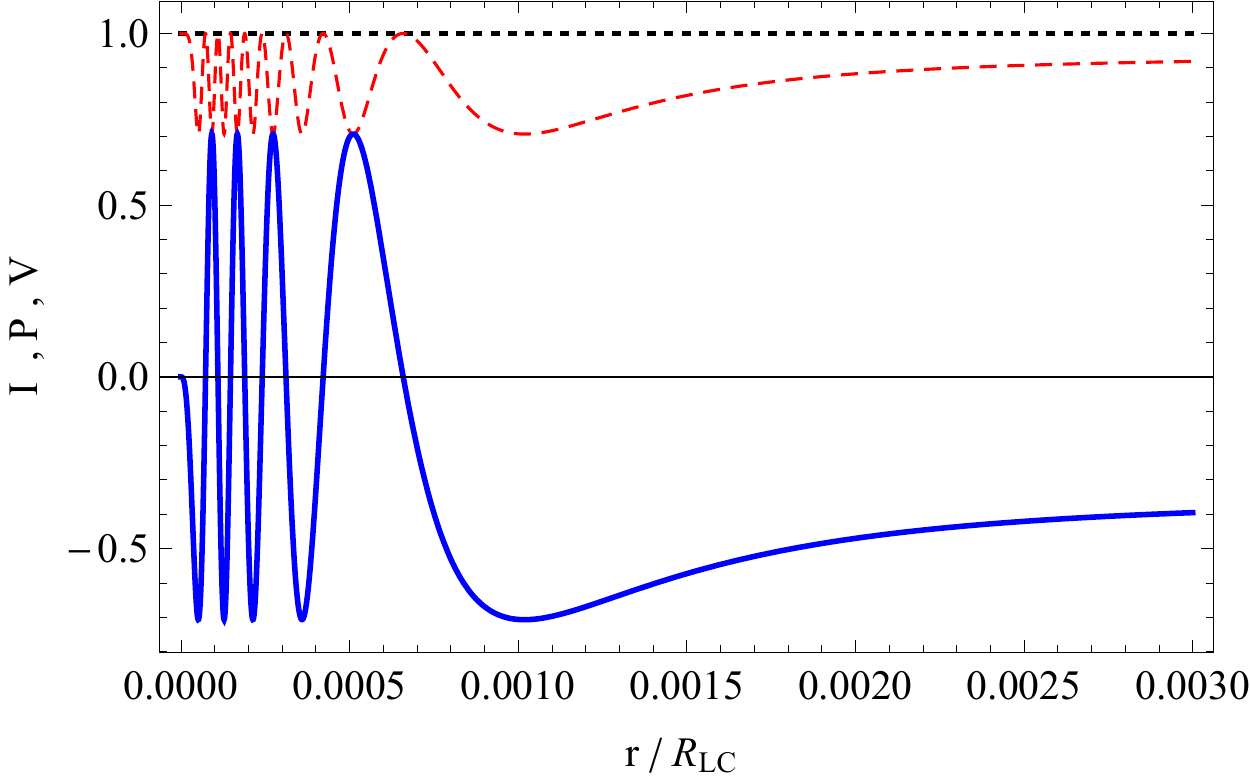}

    \caption{\label{f05} The top panels are polarization angle variation (P.A.), the middle panels are the evolution of the dimensionless  Stokes parameters U [solid (blue)] and Q [dashed (Red)] where the insert shows the variation of Q, and the lower panels are the  evolution of the dimensionless Stokes parameters I [dotted (black)], P [dashed (red)] and V [solid (blue)]. Results are shown for  a particular photon moving radially outward in the magnetosphere as a function of the dimensionless distance  $r/R_{LC}$. In this case, the parameters are the surface magnetic field $B_{0} = 10^{12} G$ (left panels), and  $B_{0} = 5\times 10^{12} G$  (right panels), inclination angle $\alpha= 15^{\circ} $, photon energy $k_{0}= 1 keV$, and rotational period $P = 1 s$.}
     \end{figure}

We also examined how the polarization position angle changes with the pulsar parameters, as are shown in Fig. \ref{f05}.  When  we solve the coupled equations given in Eq. (\ref{e42}), the Stokes parameters Q and U can also be obtained that leads to determining the position angle (P.A.) $ \psi=\frac{1}{2}\arctan({U}/{Q})$. The  variation of the amount  of P.A. depends on the magnetic field strength $B_{0}$, and  it can be increased up to $45^{\circ}$ for a given set of parameters [See Fig. \ref{f05} (right panels)]. While the QED interactions below the threshold energy of the electron-positron pair production do not modify the light amplitude in the different directions, here the vacuum rotation is a result of the spatial and temporal variations of the magnetic field. In contrast to some of the previous studies \citep{2003PhRvL..91g1101L,2007MNRAS.377.1095W,2002ApJ...566..373L,2003ApJ...588..962L} in which the resonance effect between the plasma and QED in a very specific region (resonance region) dictates several polarization characteristics of the X-ray radiation, namely (i) the conversion of photon modes following by $90^{\circ}$ jump in the position angle for the linear polarization plane; (ii) the generation of circular polarized radiation,  we have shown that nonlinear QED interaction regardless of the plasma effects does have  major effects on the polarization parameters leading to high values for the circular polarization and significant changes in the polarization angle  [See Figs. \ref{f05}].   

\section{Circular polarization of the X-ray emission from the polar cap }
\label{CAP}
Stokes parameters are additive and each photon is characterized by its own set of Stokes parameters which are defined with respect to a given frame. Therefore a polarimeter will collect a large number of seed photons to measure the polarization properties of a given source. Unlike radio or optical telescopes, which measure the intensity of the radiation from the source, most X-ray instruments detect individual photons \citep{Kislat:2015vt,Taverna:2015uu}. In the previous section, we presented the main features of polarization evolution along a single radial ray in which the magnetic dipole vector during rotation around its spin axis, intersects the line of sight (LOS) at $\omega t=0$. For concreteness, we considered X-ray emission from the open field line region around the polar cap on the star surface. The size of the cap is given by the polar cap radius $R_{pc}=R\theta_{pc}$, where $\theta_{pc}\approx\sqrt{{R}/{R_{LC}}}=\sqrt{{2\pi R}/{c P}}$ is the half-cone angle of the open field region. Several authors have considered the observed polarization properties from a small hot spot on the NS surface caused by vacuum resonance, which occurs in the dense atmospheric layers \citep{2003PhRvL..91g1101L,2009MNRAS.399.1523V}. The situation is however, different in the case of surface emission from the entire surface of NS \citep{2003MNRAS.342..134H,Taverna:2015uu}. In order to find pulsar polarization profile, it is necessary to determined the final polarization states (Stokes parameters) of the photon from each emission point on emitting surface. Stokes parameters  depend on the direction of the local magnetic field which  is in general non-uniform across the emission region, but it can be treated as uniform over a small polar cap size \citep{Taverna:2015uu}. Observed polarization direction from a surface element is correlated with the direction of the magnetic field at the polarization-limiting radius, instead of the  magnetic field direction at the emission point \citep{2003PhRvL..91g1101L}. The polarization limiting radius may occur close to either NS surface or light cylinder radius. A larger value of this radius results in a  larger polarization fraction since  the rays pass through only a  small solid angle far from the star surface, where the magnetic geometry is uniform \citep{2002PhRvD..66b3002H}. However, if the magnetic field has cylindrical symmetry,  the net circular polarization for thermal radiation coming from the whole stellar surface or a finite sized polar cap is expected to be zero in the case of  a sharp boundary for the adiabatic region \citep{2003MNRAS.342..134H,2002PhRvD..66b3002H,2003PhRvL..91g1101L,Taverna:2015uu}. According to some numerical simulations taking into account an intermediate region even in the presence of resonant cyclotron scattering, circular polarization is expected not to exceed  a few percent in the X-ray band \citep{2011ApJ...730..131F,2014MNRAS.438.1686T,Taverna:2015uu,2017MNRAS.465..492M}. Besides, it has also been shown that for rapidly rotating NSs, mode recoupling does not occur instantly at polarization-limiting radius which can results in significant circular polarization \citep{vanAdelsberg:2006uu}.

We have shown here that as the rotational frequency of the pulsar or the surface magnetic field strength increases, the polarization-limiting radius increases [ See Fig. \ref{f01} (right panels) and Fig. \ref{f04} (left panels)   ]. In our computation we have started with a  mixed state of both Q and U without any circular component V as initial condition,  we found that for some certain parameters, the generation of the high degree of circular polarization even larger than 60$\%$  is possible. It is worth mentioning that the generation of circular polarization in our model is due to conversion of linear polarization of radiation even below the adiabatic radius. In fact there is no sharp boundary for the adiabatic region in our case.

The photons that we observe from NS are emitted from different emission points, with different rotation phases which lead to the final observed Stokes parameters as a function of the rotation phase. The pulsar rotation phase can be written $\psi=\psi_{em}+\omega \Delta t=\psi_{em}+{r(t)}/{R_{LC}}$, where $\psi_{em}$ is the corresponding phase at the emission point ($\psi=0$ when B lies in the X-Z plane). Following the papers \citep{vanAdelsberg:2006uu,2010MNRAS.403..569W}, one may introduce a reference frame (X,Y,Z) in which the observer LOS is along the  \emph{Z}-axis ($\hat Z=\hat k$), and the rotation axis $\hat \Omega$ is define to be in the XZ plane. Where the angle between  $\hat \Omega$  and $\hat Z$ is denoted by $\zeta$. Furthermore, the direction of the magnetic filed B at a given point along the ray  in this frame is specified by the polar angles $(\theta$,$\phi)$ as
 \begin{align}\label{e49}
\cos\theta=\cos\zeta\cos\alpha-\sin\zeta\sin\alpha\cos\psi,
 \end{align}\label{e49}
  \begin{align}\label{e50}
\tan\phi=\frac{\sin\alpha\sin\psi}{\sin\zeta\cos\alpha+\cos\zeta\sin\alpha\cos\psi}.
 \end{align}
Where $\psi$ is the NS rotation phase and $\alpha$ is the inclination angle as we introduced before. Let us assume the photon trajectory is a straight line along  the wave vector \emph{k} with the polarization vectors ($\hat \epsilon_{X},\hat \epsilon_{Y}$)  in the $\hat X$ and $\hat Y$ directions, respectively. According to this more general configuration,  Equations (\ref{e42}) take the following forms
\begin{align}\label{e51}
\dot I=0,
\end{align}
\begin{align}\label{e52}
\dot Q=G \sin\theta^{2}\sin2\phi \ V,
\end{align}
\begin{align}\label{e53}
\dot U=-G \sin\theta^{2}\cos2\phi \ V,
\end{align}
\begin{align}\label{e54}
\dot V=-G\sin\theta^{2}\ (\sin2\phi \ Q-\cos2\phi \ U),
\end{align}
where $G$ has been defined by Eq. (\ref{e48}). It is instructive to compare this set of Eqs. (\ref{e51})-(\ref{e54}) with Eqs. (80)-(83) of \citep{vanAdelsberg:2006uu}, where instead of quantum Boltzmann equation, authors used evolution equation for the mode amplitude to drive evolution equation for the Stokes parameters.
While the phase-dependent emission from an extended area on a NS surface is computed according to different methods \citep{1983ApJ...274..846P,2002ApJ...566L..85B,2013ApJ...768..147T,2009MNRAS.399.1523V}, we stress that our main goal is not to compute X-ray polarization observables for a precise, physical model of surface emission. However, we expect that our results are not qualitatively sensitive to the polar cap size, as it has been noted in \citep{2009MNRAS.399.1523V}, the main difference between cases with various polar cap sizes are the ranges of phase over which the spot is visible. Actually, when the polarization-limiting radius lies far away from the star surface, the photons coming  from different emission points experience the same dipole field. Generalization of our computations to more realistic configurations  taking into account the  present knowledge of pulsar emission models, will be the subject of a future paper.  Furthermore,  our formalism can be applied to more complex situations when several hotspots or the entire surface of NS contribute to the X-ray emission

\section{Conclusions and Remarks}

In this paper, we have investigated the polarization properties induced in linearly polarized X-ray photons propagating through the magnetized QED vacuum of a pulsar. We solved the quantum Boltzmann equation by taking into account nonlinear photon-photon interactions within the framework of the Euler-Heisenberg Lagrangian in the presence of strong magnetic fields associated with pulsars. Consequently the set of equation describing the evolution of Stokes parameters has  obtained as presented in Eq. (\ref{e42}). With our working assumptions, we find the evolution of the Stokes parameters Q, U and V, the results are shown in Figs. \ref{f01}-\ref{f05}, where all the free parameters are varied in order to explore the polarization properties of the system. Moreover we showed that the circular component and P.A.  provide independent validations of QED vacuum polarization effects.  Although our model is based on some simple assumptions, it captures novel features of the polarization signals such as the generation of high degree of circular polarization. One of the advantages of our approach is that the quantum Boltzmann equation can be contained all possible interaction terms, giving rise source terms for the Stokes parameter V.
Measurements of the circular polarization degree are possible in the optical band \citep{2015AAS...22542101W}, while current instruments based on the photoelectric effect or Compton scattering can only measure linear polarization \citep{2014axp..book.....F}. The circular polarization fraction has been little explored so far since the low degree of circular polarization in X-ray emission of pulsars was expected. Future instruments will be able to measure both linear and circular polarization in addition to P.A. at soft X-ray energies. Our results therefore demonstrate the unique potential of X-ray polarimetry in probing the physics in the strong-field regime of astrophysical objects. Measurements of X-ray polarization, particularly when phase-resolved and measured in different energy bands give an unique opportunity to study nonlinear QED effects in the strong magnetic field of pulsars by using upcoming facilities \citep{2010NIMPA.623..766B,2010xpnw.book.....B,2013ExA....36..523S,2001Natur.411..662C,2013SPIE.8777E..0WA,2015AAS...22521401W}. It can  provide the first demonstration of QED vacuum birefringence besides many experimental efforts to detect photon-photon interaction in the ground based laboratories.

\section*{ACKNOWLEDGEMENTS}  
S. Shakeri would like to thank Prof.~R. Ruffini for supporting his visit at ICRANet Pescara, where the last part of this work was done and besides for the warm hospitality of ICRANet Faculty and Staff. He is also grateful to C. Stahl for his useful comments which substantially helped to improve this article.


\begin{thebibliography}{64}%
\makeatletter
\providecommand \@ifxundefined [1]{%
 \@ifx{#1\undefined}
}%
\providecommand \@ifnum [1]{%
 \ifnum #1\expandafter \@firstoftwo
 \else \expandafter \@secondoftwo
 \fi
}%
\providecommand \@ifx [1]{%
 \ifx #1\expandafter \@firstoftwo
 \else \expandafter \@secondoftwo
 \fi
}%
\providecommand \natexlab [1]{#1}%
\providecommand \enquote  [1]{``#1''}%
\providecommand \bibnamefont  [1]{#1}%
\providecommand \bibfnamefont [1]{#1}%
\providecommand \citenamefont [1]{#1}%
\providecommand \href@noop [0]{\@secondoftwo}%
\providecommand \href [0]{\begingroup \@sanitize@url \@href}%
\providecommand \@href[1]{\@@startlink{#1}\@@href}%
\providecommand \@@href[1]{\endgroup#1\@@endlink}%
\providecommand \@sanitize@url [0]{\catcode `\\12\catcode `\$12\catcode
  `\&12\catcode `\#12\catcode `\^12\catcode `\_12\catcode `\%12\relax}%
\providecommand \@@startlink[1]{}%
\providecommand \@@endlink[0]{}%
\providecommand \url  [0]{\begingroup\@sanitize@url \@url }%
\providecommand \@url [1]{\endgroup\@href {#1}{\urlprefix }}%
\providecommand \urlprefix  [0]{URL }%
\providecommand \Eprint [0]{\href }%
\providecommand \doibase [0]{http://dx.doi.org/}%
\providecommand \selectlanguage [0]{\@gobble}%
\providecommand \bibinfo  [0]{\@secondoftwo}%
\providecommand \bibfield  [0]{\@secondoftwo}%
\providecommand \translation [1]{[#1]}%
\providecommand \BibitemOpen [0]{}%
\providecommand \bibitemStop [0]{}%
\providecommand \bibitemNoStop [0]{.\EOS\space}%
\providecommand \EOS [0]{\spacefactor3000\relax}%
\providecommand \BibitemShut  [1]{\csname bibitem#1\endcsname}%
\let\auto@bib@innerbib\@empty
\bibitem [{\citenamefont {Euler}\ and\ \citenamefont
  {Kockel}(1935)}]{1935NW.....23..246E}%
  \BibitemOpen
  \bibfield  {author} {\bibinfo {author} {\bibfnamefont {H.}~\bibnamefont
  {Euler}}\ and\ \bibinfo {author} {\bibfnamefont {B.}~\bibnamefont {Kockel}},\
  }\href@noop {} {\bibfield  {journal} {\bibinfo  {journal} {Die
  Naturwissenschaften}\ }\textbf {\bibinfo {volume} {23}},\ \bibinfo {pages}
  {246} (\bibinfo {year} {1935})}\BibitemShut {NoStop}%
\bibitem [{\citenamefont {Heisenberg}\ and\ \citenamefont
  {Euler}(1936)}]{1936ZPhy...98..714H}%
  \BibitemOpen
  \bibfield  {author} {\bibinfo {author} {\bibfnamefont {W.}~\bibnamefont
  {Heisenberg}}\ and\ \bibinfo {author} {\bibfnamefont {H.}~\bibnamefont
  {Euler}},\ }\href@noop {} {\bibfield  {journal} {\bibinfo  {journal}
  {Zeitschrift f{\"u}r Physik}\ }\textbf {\bibinfo {volume} {98}},\ \bibinfo
  {pages} {714} (\bibinfo {year} {1936})}\BibitemShut {NoStop}%
\bibitem [{\citenamefont {Adler}(1971)}]{1971AnPhy..67..599A}%
  \BibitemOpen
  \bibfield  {author} {\bibinfo {author} {\bibfnamefont {S.~L.}\ \bibnamefont
  {Adler}},\ }\href@noop {} {\bibfield  {journal} {\bibinfo  {journal} {Ann.
  Phys. (USA)}\ }\textbf {\bibinfo {volume} {67}},\ \bibinfo {pages} {599}
  (\bibinfo {year} {1971})}\BibitemShut {NoStop}%
\bibitem [{\citenamefont {Heyl}\ and\ \citenamefont
  {Hernquist}(1997)}]{1997JPhA...30.6485H}%
  \BibitemOpen
  \bibfield  {author} {\bibinfo {author} {\bibfnamefont {J.~S.}\ \bibnamefont
  {Heyl}}\ and\ \bibinfo {author} {\bibfnamefont {L.}~\bibnamefont
  {Hernquist}},\ }\href@noop {} {\bibfield  {journal} {\bibinfo  {journal}
  {Journal of Physics A: Mathematical and General}\ }\textbf {\bibinfo {volume}
  {30}},\ \bibinfo {pages} {6485} (\bibinfo {year} {1997})}\BibitemShut
  {NoStop}%
\bibitem [{\citenamefont {di~Piazza}\ \emph {et~al.}(2012)\citenamefont
  {di~Piazza}, \citenamefont {M{\"u}ller}, \citenamefont {Hatsagortsyan},\ and\
  \citenamefont {Keitel}}]{2012RvMP...84.1177D}%
  \BibitemOpen
  \bibfield  {author} {\bibinfo {author} {\bibfnamefont {A.}~\bibnamefont
  {di~Piazza}}, \bibinfo {author} {\bibfnamefont {C.}~\bibnamefont
  {M{\"u}ller}}, \bibinfo {author} {\bibfnamefont {K.~Z.}\ \bibnamefont
  {Hatsagortsyan}}, \ and\ \bibinfo {author} {\bibfnamefont {C.~H.}\
  \bibnamefont {Keitel}},\ }\href@noop {} {\bibfield  {journal} {\bibinfo
  {journal} {Reviews of Modern Physics}\ }\textbf {\bibinfo {volume} {84}},\
  \bibinfo {pages} {1177} (\bibinfo {year} {2012})}\BibitemShut {NoStop}%
\bibitem [{\citenamefont {Schlenvoigt}\ \emph {et~al.}(2016)\citenamefont
  {Schlenvoigt}, \citenamefont {Heinzl}, \citenamefont {Schramm}, \citenamefont
  {Cowan},\ and\ \citenamefont {Sauerbrey}}]{2016PhyS...91b3010S}%
  \BibitemOpen
  \bibfield  {author} {\bibinfo {author} {\bibfnamefont {H.-P.}\ \bibnamefont
  {Schlenvoigt}}, \bibinfo {author} {\bibfnamefont {T.}~\bibnamefont {Heinzl}},
  \bibinfo {author} {\bibfnamefont {U.}~\bibnamefont {Schramm}}, \bibinfo
  {author} {\bibfnamefont {T.~E.}\ \bibnamefont {Cowan}}, \ and\ \bibinfo
  {author} {\bibfnamefont {R.}~\bibnamefont {Sauerbrey}},\ }\href@noop {}
  {\bibfield  {journal} {\bibinfo  {journal} {Physica Scripta}\ }\textbf
  {\bibinfo {volume} {91}},\ \bibinfo {pages} {023010} (\bibinfo {year}
  {2016})}\BibitemShut {NoStop}%
\bibitem [{HIB()}]{HIBEF}%
  \BibitemOpen
  \href@noop {} {\emph {\bibinfo {title} {{HIBEF website}}}},\ \bibinfo {note}
  {\url{http://www.hzdr.de/hibef}}\BibitemShut {NoStop}%
\bibitem [{ELI()}]{ELI}%
  \BibitemOpen
  \href@noop {} {\emph {\bibinfo {title} {{ELI}}}},\ \bibinfo {note}
  {\url{http://www.hzdr.de/hibef, "HiPER", http://www.hiperlaser.org, "XCELS",
  http://www.xcels.iapras.ru/img/XCELS-Project-english-version.pdf}}\BibitemShut
  {NoStop}%
\bibitem [{\citenamefont {Ventura}\ \emph {et~al.}(1979)\citenamefont
  {Ventura}, \citenamefont {Nagel},\ and\ \citenamefont
  {Meszaros}}]{Ventura:1979gi}%
  \BibitemOpen
  \bibfield  {author} {\bibinfo {author} {\bibfnamefont {J.}~\bibnamefont
  {Ventura}}, \bibinfo {author} {\bibfnamefont {W.}~\bibnamefont {Nagel}}, \
  and\ \bibinfo {author} {\bibfnamefont {P.}~\bibnamefont {Meszaros}},\
  }\href@noop {} {\bibfield  {journal} {\bibinfo  {journal} {Astrophys. J.}\
  }\textbf {\bibinfo {volume} {233}},\ \bibinfo {pages} {L125} (\bibinfo {year}
  {1979})}\BibitemShut {NoStop}%
\bibitem [{\citenamefont {van Adelsberg}\ and\ \citenamefont
  {Lai}(2006)}]{vanAdelsberg:2006uu}%
  \BibitemOpen
  \bibfield  {author} {\bibinfo {author} {\bibfnamefont {M.}~\bibnamefont {van
  Adelsberg}}\ and\ \bibinfo {author} {\bibfnamefont {D.}~\bibnamefont {Lai}},\
  }\href@noop {} {\bibfield  {journal} {\bibinfo  {journal} {Mon. Not. Roy.
  Astron. Soc.}\ }\textbf {\bibinfo {volume} {373}},\ \bibinfo {pages} {1495}
  (\bibinfo {year} {2006})}\BibitemShut {NoStop}%
\bibitem [{\citenamefont {Fern{\'a}ndez}\ and\ \citenamefont
  {Davis}(2011)}]{2011ApJ...730..131F}%
  \BibitemOpen
  \bibfield  {author} {\bibinfo {author} {\bibfnamefont {R.}~\bibnamefont
  {Fern{\'a}ndez}}\ and\ \bibinfo {author} {\bibfnamefont {S.~W.}\ \bibnamefont
  {Davis}},\ }\href@noop {} {\bibfield  {journal} {\bibinfo  {journal} {The
  Astrophysical Journal}\ }\textbf {\bibinfo {volume} {730}},\ \bibinfo {pages}
  {131} (\bibinfo {year} {2011})}\BibitemShut {NoStop}%
\bibitem [{\citenamefont {{Novick}}\ \emph {et~al.}(1977)\citenamefont
  {{Novick}}, \citenamefont {{Weisskopf}}, \citenamefont {{Angel}},\ and\
  \citenamefont {{Sutherland}}}]{1977ApJ...215L.117N}%
  \BibitemOpen
  \bibfield  {author} {\bibinfo {author} {\bibfnamefont {R.}~\bibnamefont
  {{Novick}}}, \bibinfo {author} {\bibfnamefont {M.~C.}\ \bibnamefont
  {{Weisskopf}}}, \bibinfo {author} {\bibfnamefont {J.~R.~P.}\ \bibnamefont
  {{Angel}}}, \ and\ \bibinfo {author} {\bibfnamefont {P.~G.}\ \bibnamefont
  {{Sutherland}}},\ }\href {\doibase 10.1086/182492} {\bibfield  {journal}
  {\bibinfo  {journal} {Astrophysical Journal}\ }\textbf {\bibinfo {volume}
  {215}},\ \bibinfo {pages} {L117} (\bibinfo {year} {1977})}\BibitemShut
  {NoStop}%
\bibitem [{\citenamefont {Meszaros}\ and\ \citenamefont
  {Ventura}(1979)}]{1979PhRvD..19.3565M}%
  \BibitemOpen
  \bibfield  {author} {\bibinfo {author} {\bibfnamefont {P.}~\bibnamefont
  {Meszaros}}\ and\ \bibinfo {author} {\bibfnamefont {J.}~\bibnamefont
  {Ventura}},\ }\href@noop {} {\bibfield  {journal} {\bibinfo  {journal}
  {Physical Review D - Particles and Fields}\ }\textbf {\bibinfo {volume}
  {19}},\ \bibinfo {pages} {3565} (\bibinfo {year} {1979})}\BibitemShut
  {NoStop}%
\bibitem [{\citenamefont {Lai}\ and\ \citenamefont
  {Ho}(2003{\natexlab{a}})}]{2003PhRvL..91g1101L}%
  \BibitemOpen
  \bibfield  {author} {\bibinfo {author} {\bibfnamefont {D.}~\bibnamefont
  {Lai}}\ and\ \bibinfo {author} {\bibfnamefont {W.~C.}\ \bibnamefont {Ho}},\
  }\href@noop {} {\bibfield  {journal} {\bibinfo  {journal} {Physical Review
  Letters}\ }\textbf {\bibinfo {volume} {91}},\ \bibinfo {pages} {071101}
  (\bibinfo {year} {2003}{\natexlab{a}})}\BibitemShut {NoStop}%
\bibitem [{\citenamefont {Wang}\ and\ \citenamefont
  {Lai}(2007)}]{2007MNRAS.377.1095W}%
  \BibitemOpen
  \bibfield  {author} {\bibinfo {author} {\bibfnamefont {C.}~\bibnamefont
  {Wang}}\ and\ \bibinfo {author} {\bibfnamefont {D.}~\bibnamefont {Lai}},\
  }\href@noop {} {\bibfield  {journal} {\bibinfo  {journal} {Monthly Notices of
  the Royal Astronomical Society}\ }\textbf {\bibinfo {volume} {377}},\
  \bibinfo {pages} {1095} (\bibinfo {year} {2007})}\BibitemShut {NoStop}%
\bibitem [{\citenamefont {Shannon}\ and\ \citenamefont
  {Heyl}(2006)}]{2006MNRAS.368.1377S}%
  \BibitemOpen
  \bibfield  {author} {\bibinfo {author} {\bibfnamefont {R.~M.}\ \bibnamefont
  {Shannon}}\ and\ \bibinfo {author} {\bibfnamefont {J.~S.}\ \bibnamefont
  {Heyl}},\ }\href@noop {} {\bibfield  {journal} {\bibinfo  {journal} {Monthly
  Notices of the Royal Astronomical Society}\ }\textbf {\bibinfo {volume}
  {368}},\ \bibinfo {pages} {1377} (\bibinfo {year} {2006})}\BibitemShut
  {NoStop}%
\bibitem [{\citenamefont {Wang}\ \emph {et~al.}(2010)\citenamefont {Wang},
  \citenamefont {Lai},\ and\ \citenamefont {Han}}]{2010MNRAS.403..569W}%
  \BibitemOpen
  \bibfield  {author} {\bibinfo {author} {\bibfnamefont {C.}~\bibnamefont
  {Wang}}, \bibinfo {author} {\bibfnamefont {D.}~\bibnamefont {Lai}}, \ and\
  \bibinfo {author} {\bibfnamefont {J.}~\bibnamefont {Han}},\ }\href@noop {}
  {\bibfield  {journal} {\bibinfo  {journal} {Monthly Notices of the Royal
  Astronomical Society}\ }\textbf {\bibinfo {volume} {403}},\ \bibinfo {pages}
  {569} (\bibinfo {year} {2010})}\BibitemShut {NoStop}%
\bibitem [{\citenamefont {{Lai, Dong}}\ and\ \citenamefont {{Ho, Wynn C
  G}}(2002)}]{2002ApJ...566..373L}%
  \BibitemOpen
  \bibfield  {author} {\bibinfo {author} {\bibnamefont {{Lai, Dong}}}\ and\
  \bibinfo {author} {\bibnamefont {{Ho, Wynn C G}}},\ }\href@noop {} {\bibfield
   {journal} {\bibinfo  {journal} {The Astrophysical Journal}\ }\textbf
  {\bibinfo {volume} {566}},\ \bibinfo {pages} {373} (\bibinfo {year}
  {2002})}\BibitemShut {NoStop}%
\bibitem [{\citenamefont {Lai}\ and\ \citenamefont
  {Ho}(2003{\natexlab{b}})}]{2003ApJ...588..962L}%
  \BibitemOpen
  \bibfield  {author} {\bibinfo {author} {\bibfnamefont {D.}~\bibnamefont
  {Lai}}\ and\ \bibinfo {author} {\bibfnamefont {W.~C.~G.}\ \bibnamefont
  {Ho}},\ }\href {\doibase 10.1086/374334} {\bibfield  {journal} {\bibinfo
  {journal} {Astrophys. J.}\ }\textbf {\bibinfo {volume} {588}},\ \bibinfo
  {pages} {962} (\bibinfo {year} {2003}{\natexlab{b}})},\ \Eprint
  {http://arxiv.org/abs/astro-ph/0211315} {arXiv:astro-ph/0211315 [astro-ph]}
  \BibitemShut {NoStop}%
\bibitem [{\citenamefont {Wang}\ and\ \citenamefont
  {Lai}(2009)}]{2009MNRAS.398..515W}%
  \BibitemOpen
  \bibfield  {author} {\bibinfo {author} {\bibfnamefont {C.}~\bibnamefont
  {Wang}}\ and\ \bibinfo {author} {\bibfnamefont {D.}~\bibnamefont {Lai}},\
  }\href@noop {} {\bibfield  {journal} {\bibinfo  {journal} {Monthly Notices of
  the Royal Astronomical Society}\ }\textbf {\bibinfo {volume} {398}},\
  \bibinfo {pages} {515} (\bibinfo {year} {2009})}\BibitemShut {NoStop}%
\bibitem [{\citenamefont {Pavlov}\ and\ \citenamefont
  {Shibanov}(1978)}]{1978SvA....22..214P}%
  \BibitemOpen
  \bibfield  {author} {\bibinfo {author} {\bibfnamefont {G.~G.}\ \bibnamefont
  {Pavlov}}\ and\ \bibinfo {author} {\bibfnamefont {I.~A.}\ \bibnamefont
  {Shibanov}},\ }\href@noop {} {\bibfield  {journal} {\bibinfo  {journal}
  {Soviet Astronomy}\ }\textbf {\bibinfo {volume} {22}},\ \bibinfo {pages}
  {214} (\bibinfo {year} {1978})}\BibitemShut {NoStop}%
\bibitem [{\citenamefont {Harding}\ and\ \citenamefont
  {Lai}(2006)}]{2006RPPh...69.2631H}%
  \BibitemOpen
  \bibfield  {author} {\bibinfo {author} {\bibfnamefont {A.~K.}\ \bibnamefont
  {Harding}}\ and\ \bibinfo {author} {\bibfnamefont {D.}~\bibnamefont {Lai}},\
  }\href@noop {} {\bibfield  {journal} {\bibinfo  {journal} {Reports on
  Progress in Physics}\ }\textbf {\bibinfo {volume} {69}},\ \bibinfo {pages}
  {2631} (\bibinfo {year} {2006})}\BibitemShut {NoStop}%
\bibitem [{\citenamefont {Pavlov}\ and\ \citenamefont
  {Zavlin}(2000)}]{2000ApJ...529.1011P}%
  \BibitemOpen
  \bibfield  {author} {\bibinfo {author} {\bibfnamefont {G.~G.}\ \bibnamefont
  {Pavlov}}\ and\ \bibinfo {author} {\bibfnamefont {V.~E.}\ \bibnamefont
  {Zavlin}},\ }\href@noop {} {\bibfield  {journal} {\bibinfo  {journal} {The
  Astrophysical Journal}\ }\textbf {\bibinfo {volume} {529}},\ \bibinfo {pages}
  {1011} (\bibinfo {year} {2000})}\BibitemShut {NoStop}%
\bibitem [{\citenamefont {Heyl}\ and\ \citenamefont
  {Shaviv}(2000)}]{2000MNRAS.311..555H}%
  \BibitemOpen
  \bibfield  {author} {\bibinfo {author} {\bibfnamefont {J.~S.}\ \bibnamefont
  {Heyl}}\ and\ \bibinfo {author} {\bibfnamefont {N.~J.}\ \bibnamefont
  {Shaviv}},\ }\href@noop {} {\bibfield  {journal} {\bibinfo  {journal}
  {Monthly Notices of the Royal Astronomical Society}\ }\textbf {\bibinfo
  {volume} {311}},\ \bibinfo {pages} {555} (\bibinfo {year}
  {2000})}\BibitemShut {NoStop}%
\bibitem [{\citenamefont {Heyl}\ and\ \citenamefont
  {Shaviv}(2002)}]{2002PhRvD..66b3002H}%
  \BibitemOpen
  \bibfield  {author} {\bibinfo {author} {\bibfnamefont {J.~S.}\ \bibnamefont
  {Heyl}}\ and\ \bibinfo {author} {\bibfnamefont {N.~J.}\ \bibnamefont
  {Shaviv}},\ }\href@noop {} {\bibfield  {journal} {\bibinfo  {journal}
  {Physical Review D}\ }\textbf {\bibinfo {volume} {66}},\ \bibinfo {pages}
  {023002} (\bibinfo {year} {2002})}\BibitemShut {NoStop}%
\bibitem [{\citenamefont {Heyl}\ \emph {et~al.}(2003)\citenamefont {Heyl},
  \citenamefont {Shaviv},\ and\ \citenamefont {Lloyd}}]{2003MNRAS.342..134H}%
  \BibitemOpen
  \bibfield  {author} {\bibinfo {author} {\bibfnamefont {J.~S.}\ \bibnamefont
  {Heyl}}, \bibinfo {author} {\bibfnamefont {N.~J.}\ \bibnamefont {Shaviv}}, \
  and\ \bibinfo {author} {\bibfnamefont {D.}~\bibnamefont {Lloyd}},\
  }\href@noop {} {\bibfield  {journal} {\bibinfo  {journal} {Monthly Notice of
  the Royal Astronomical Society}\ }\textbf {\bibinfo {volume} {342}},\
  \bibinfo {pages} {134} (\bibinfo {year} {2003})}\BibitemShut {NoStop}%
\bibitem [{\citenamefont {Mignani}\ \emph {et~al.}(2017)\citenamefont
  {Mignani}, \citenamefont {Testa}, \citenamefont {Gonz{\'a}lez~Caniulef},
  \citenamefont {Taverna}, \citenamefont {Turolla}, \citenamefont {Zane},\ and\
  \citenamefont {Wu}}]{2017MNRAS.465..492M}%
  \BibitemOpen
  \bibfield  {author} {\bibinfo {author} {\bibfnamefont {R.~P.}\ \bibnamefont
  {Mignani}}, \bibinfo {author} {\bibfnamefont {V.}~\bibnamefont {Testa}},
  \bibinfo {author} {\bibfnamefont {D.}~\bibnamefont {Gonz{\'a}lez~Caniulef}},
  \bibinfo {author} {\bibfnamefont {R.}~\bibnamefont {Taverna}}, \bibinfo
  {author} {\bibfnamefont {R.}~\bibnamefont {Turolla}}, \bibinfo {author}
  {\bibfnamefont {S.}~\bibnamefont {Zane}}, \ and\ \bibinfo {author}
  {\bibfnamefont {K.}~\bibnamefont {Wu}},\ }\href@noop {} {\bibfield  {journal}
  {\bibinfo  {journal} {Monthly Notices of the Royal Astronomical Society}\
  }\textbf {\bibinfo {volume} {465}},\ \bibinfo {pages} {492} (\bibinfo {year}
  {2017})}\BibitemShut {NoStop}%
\bibitem [{\citenamefont {Caniulef}\ \emph {et~al.}(2016)\citenamefont
  {Caniulef}, \citenamefont {Zane}, \citenamefont {Taverna}, \citenamefont
  {Turolla},\ and\ \citenamefont {Wu}}]{Caniulef:2016vb}%
  \BibitemOpen
  \bibfield  {author} {\bibinfo {author} {\bibfnamefont {D.~G.}\ \bibnamefont
  {Caniulef}}, \bibinfo {author} {\bibfnamefont {S.}~\bibnamefont {Zane}},
  \bibinfo {author} {\bibfnamefont {R.}~\bibnamefont {Taverna}}, \bibinfo
  {author} {\bibfnamefont {R.}~\bibnamefont {Turolla}}, \ and\ \bibinfo
  {author} {\bibfnamefont {K.}~\bibnamefont {Wu}},\ }\href@noop {} {\bibfield
  {journal} {\bibinfo  {journal} {Monthly Notices of the Royal Astronomical
  Society}\ }\textbf {\bibinfo {volume} {459}},\ \bibinfo {pages} {3585}
  (\bibinfo {year} {2016})}\BibitemShut {NoStop}%
\bibitem [{\citenamefont {Walter}\ and\ \citenamefont
  {Matthews}(1997)}]{1997Natur.389..358W}%
  \BibitemOpen
  \bibfield  {author} {\bibinfo {author} {\bibfnamefont {F.~M.}\ \bibnamefont
  {Walter}}\ and\ \bibinfo {author} {\bibfnamefont {L.~D.}\ \bibnamefont
  {Matthews}},\ }\href@noop {} {\bibfield  {journal} {\bibinfo  {journal}
  {Nature}\ }\textbf {\bibinfo {volume} {389}},\ \bibinfo {pages} {358}
  (\bibinfo {year} {1997})}\BibitemShut {NoStop}%
\bibitem [{\citenamefont {Mignani}\ \emph {et~al.}(2013)\citenamefont
  {Mignani}, \citenamefont {Vande~Putte}, \citenamefont {Cropper},
  \citenamefont {Turolla}, \citenamefont {Zane}, \citenamefont {Pellizza},
  \citenamefont {Bignone}, \citenamefont {Sartore},\ and\ \citenamefont
  {Treves}}]{2013MNRAS.429.3517M}%
  \BibitemOpen
  \bibfield  {author} {\bibinfo {author} {\bibfnamefont {R.~P.}\ \bibnamefont
  {Mignani}}, \bibinfo {author} {\bibfnamefont {D.}~\bibnamefont
  {Vande~Putte}}, \bibinfo {author} {\bibfnamefont {M.}~\bibnamefont
  {Cropper}}, \bibinfo {author} {\bibfnamefont {R.}~\bibnamefont {Turolla}},
  \bibinfo {author} {\bibfnamefont {S.}~\bibnamefont {Zane}}, \bibinfo {author}
  {\bibfnamefont {L.~J.}\ \bibnamefont {Pellizza}}, \bibinfo {author}
  {\bibfnamefont {L.~A.}\ \bibnamefont {Bignone}}, \bibinfo {author}
  {\bibfnamefont {N.}~\bibnamefont {Sartore}}, \ and\ \bibinfo {author}
  {\bibfnamefont {A.}~\bibnamefont {Treves}},\ }\href@noop {} {\bibfield
  {journal} {\bibinfo  {journal} {Monthly Notices of the Royal Astronomical
  Society}\ }\textbf {\bibinfo {volume} {429}},\ \bibinfo {pages} {3517}
  (\bibinfo {year} {2013})}\BibitemShut {NoStop}%
\bibitem [{\citenamefont {Novick}\ \emph {et~al.}(1972)\citenamefont {Novick},
  \citenamefont {Weisskopf}, \citenamefont {Berthelsdorf}, \citenamefont
  {Linke},\ and\ \citenamefont {Wolff}}]{1972ApJ...174L...1N}%
  \BibitemOpen
  \bibfield  {author} {\bibinfo {author} {\bibfnamefont {R.}~\bibnamefont
  {Novick}}, \bibinfo {author} {\bibfnamefont {M.~C.}\ \bibnamefont
  {Weisskopf}}, \bibinfo {author} {\bibfnamefont {R.}~\bibnamefont
  {Berthelsdorf}}, \bibinfo {author} {\bibfnamefont {R.}~\bibnamefont {Linke}},
  \ and\ \bibinfo {author} {\bibfnamefont {R.~S.}\ \bibnamefont {Wolff}},\
  }\href@noop {} {\bibfield  {journal} {\bibinfo  {journal} {Astrophysical
  Journal}\ }\textbf {\bibinfo {volume} {174}},\ \bibinfo {pages} {L1}
  (\bibinfo {year} {1972})}\BibitemShut {NoStop}%
\bibitem [{\citenamefont {Weisskopf}\ \emph {et~al.}(1978)\citenamefont
  {Weisskopf}, \citenamefont {Silver}, \citenamefont {Kestenbaum},
  \citenamefont {Long},\ and\ \citenamefont {Novick}}]{1978ApJ...220L.117W}%
  \BibitemOpen
  \bibfield  {author} {\bibinfo {author} {\bibfnamefont {M.~C.}\ \bibnamefont
  {Weisskopf}}, \bibinfo {author} {\bibfnamefont {E.~H.}\ \bibnamefont
  {Silver}}, \bibinfo {author} {\bibfnamefont {H.~L.}\ \bibnamefont
  {Kestenbaum}}, \bibinfo {author} {\bibfnamefont {K.~S.}\ \bibnamefont
  {Long}}, \ and\ \bibinfo {author} {\bibfnamefont {R.}~\bibnamefont
  {Novick}},\ }\href@noop {} {\bibfield  {journal} {\bibinfo  {journal}
  {Astrophysical Journal}\ }\textbf {\bibinfo {volume} {220}},\ \bibinfo
  {pages} {L117} (\bibinfo {year} {1978})}\BibitemShut {NoStop}%
\bibitem [{\citenamefont {Bellazzini}\ and\ \citenamefont
  {Muleri}(2010)}]{2010NIMPA.623..766B}%
  \BibitemOpen
  \bibfield  {author} {\bibinfo {author} {\bibfnamefont {R.}~\bibnamefont
  {Bellazzini}}\ and\ \bibinfo {author} {\bibfnamefont {F.}~\bibnamefont
  {Muleri}},\ }\href@noop {} {\bibfield  {journal} {\bibinfo  {journal}
  {Nuclear Instruments and Methods in Physics Research Section A}\ }\textbf
  {\bibinfo {volume} {623}},\ \bibinfo {pages} {766} (\bibinfo {year}
  {2010})}\BibitemShut {NoStop}%
\bibitem [{\citenamefont {Bellazzini}\ \emph {et~al.}(2010)\citenamefont
  {Bellazzini}, \citenamefont {Costa}, \citenamefont {Matt},\ and\
  \citenamefont {Tagliaferri}}]{2010xpnw.book.....B}%
  \BibitemOpen
  \bibfield  {author} {\bibinfo {author} {\bibfnamefont {R.}~\bibnamefont
  {Bellazzini}}, \bibinfo {author} {\bibfnamefont {E.}~\bibnamefont {Costa}},
  \bibinfo {author} {\bibfnamefont {G.}~\bibnamefont {Matt}}, \ and\ \bibinfo
  {author} {\bibfnamefont {G.}~\bibnamefont {Tagliaferri}},\ }\href@noop {}
  {\bibfield  {journal} {\bibinfo  {journal} {X-ray Polarimetry: A New Window
  in Astrophysics by Ronaldo Bellazzini}\ } (\bibinfo {year}
  {2010})}\BibitemShut {NoStop}%
\bibitem [{\citenamefont {Costa}\ \emph {et~al.}(2001)\citenamefont {Costa},
  \citenamefont {Soffitta}, \citenamefont {Bellazzini}, \citenamefont {Brez},
  \citenamefont {Lumb},\ and\ \citenamefont {Spandre}}]{2001Natur.411..662C}%
  \BibitemOpen
  \bibfield  {author} {\bibinfo {author} {\bibfnamefont {E.}~\bibnamefont
  {Costa}}, \bibinfo {author} {\bibfnamefont {P.}~\bibnamefont {Soffitta}},
  \bibinfo {author} {\bibfnamefont {R.}~\bibnamefont {Bellazzini}}, \bibinfo
  {author} {\bibfnamefont {A.}~\bibnamefont {Brez}}, \bibinfo {author}
  {\bibfnamefont {N.}~\bibnamefont {Lumb}}, \ and\ \bibinfo {author}
  {\bibfnamefont {G.}~\bibnamefont {Spandre}},\ }\href@noop {} {\bibfield
  {journal} {\bibinfo  {journal} {Nature}\ }\textbf {\bibinfo {volume} {411}},\
  \bibinfo {pages} {662} (\bibinfo {year} {2001})}\BibitemShut {NoStop}%
\bibitem [{\citenamefont {Soffitta}\ \emph {et~al.}(2013)\citenamefont
  {Soffitta}, \citenamefont {Barcons}, \citenamefont {Bellazzini},
  \citenamefont {Braga}, \citenamefont {Costa}, \citenamefont {Fraser},
  \citenamefont {Gburek}, \citenamefont {Huovelin}, \citenamefont {Matt},
  \citenamefont {Pearce}, \citenamefont {Poutanen}, \citenamefont {Reglero},
  \citenamefont {Santangelo}, \citenamefont {Sunyaev}, \citenamefont
  {Tagliaferri}, \citenamefont {Weisskopf}, \citenamefont {Aloisio},
  \citenamefont {Amato}, \citenamefont {Attin{\'a}}, \citenamefont {Axelsson},
  \citenamefont {Baldini}, \citenamefont {Basso}, \citenamefont {Bianchi},
  \citenamefont {Blasi}, \citenamefont {Bregeon}, \citenamefont {Brez},
  \citenamefont {Bucciantini}, \citenamefont {Burderi}, \citenamefont
  {Burwitz}, \citenamefont {Casella}, \citenamefont {Churazov}, \citenamefont
  {Civitani}, \citenamefont {Covino}, \citenamefont {Curado~da Silva},
  \citenamefont {Cusumano}, \citenamefont {Dadina}, \citenamefont {D'Amico},
  \citenamefont {De~Rosa}, \citenamefont {Di~Cosimo}, \citenamefont
  {Di~Persio}, \citenamefont {Di~Salvo}, \citenamefont {Dovciak}, \citenamefont
  {Elsner}, \citenamefont {Eyles}, \citenamefont {Fabian}, \citenamefont
  {Fabiani}, \citenamefont {Feng}, \citenamefont {Giarrusso}, \citenamefont
  {Goosmann}, \citenamefont {Grandi}, \citenamefont {Grosso}, \citenamefont
  {Israel}, \citenamefont {Jackson}, \citenamefont {Kaaret}, \citenamefont
  {Karas}, \citenamefont {Kuss}, \citenamefont {Lai}, \citenamefont {Rosa},
  \citenamefont {Larsson}, \citenamefont {Larsson}, \citenamefont {Latronico},
  \citenamefont {Maggio}, \citenamefont {Maia}, \citenamefont {Marin},
  \citenamefont {Massai}, \citenamefont {Mineo}, \citenamefont {Minuti},
  \citenamefont {Moretti}, \citenamefont {Muleri}, \citenamefont {O'Dell},
  \citenamefont {Pareschi}, \citenamefont {Peres}, \citenamefont {Pesce},
  \citenamefont {Petrucci}, \citenamefont {Pinchera}, \citenamefont {Porquet},
  \citenamefont {Ramsey}, \citenamefont {Rea}, \citenamefont {Reale},
  \citenamefont {Rodrigo}, \citenamefont {R{\'o}{\.{z}}a{\'{n}}ska},
  \citenamefont {Rubini}, \citenamefont {Rudawy}, \citenamefont {Ryde},
  \citenamefont {Salvati}, \citenamefont {de~Santiago}, \citenamefont
  {Sazonov}, \citenamefont {Sgr{\'o}}, \citenamefont {Silver}, \citenamefont
  {Spandre}, \citenamefont {Spiga}, \citenamefont {Stella}, \citenamefont
  {Tamagawa}, \citenamefont {Tamborra}, \citenamefont {Tavecchio},
  \citenamefont {Teixeira~Dias}, \citenamefont {van Adelsberg}, \citenamefont
  {Wu},\ and\ \citenamefont {Zane}}]{2013ExA....36..523S}%
  \BibitemOpen
  \bibfield  {author} {\bibinfo {author} {\bibfnamefont {P.}~\bibnamefont
  {Soffitta}}, \bibinfo {author} {\bibfnamefont {X.}~\bibnamefont {Barcons}},
  \bibinfo {author} {\bibfnamefont {R.}~\bibnamefont {Bellazzini}}, \bibinfo
  {author} {\bibfnamefont {J.}~\bibnamefont {Braga}}, \bibinfo {author}
  {\bibfnamefont {E.}~\bibnamefont {Costa}}, \bibinfo {author} {\bibfnamefont
  {G.~W.}\ \bibnamefont {Fraser}}, \bibinfo {author} {\bibfnamefont
  {S.}~\bibnamefont {Gburek}}, \bibinfo {author} {\bibfnamefont
  {J.}~\bibnamefont {Huovelin}}, \bibinfo {author} {\bibfnamefont
  {G.}~\bibnamefont {Matt}}, \bibinfo {author} {\bibfnamefont {M.}~\bibnamefont
  {Pearce}}, \bibinfo {author} {\bibfnamefont {J.}~\bibnamefont {Poutanen}},
  \bibinfo {author} {\bibfnamefont {V.}~\bibnamefont {Reglero}}, \bibinfo
  {author} {\bibfnamefont {A.}~\bibnamefont {Santangelo}}, \bibinfo {author}
  {\bibfnamefont {R.~A.}\ \bibnamefont {Sunyaev}}, \bibinfo {author}
  {\bibfnamefont {G.}~\bibnamefont {Tagliaferri}}, \bibinfo {author}
  {\bibfnamefont {M.}~\bibnamefont {Weisskopf}}, \bibinfo {author}
  {\bibfnamefont {R.}~\bibnamefont {Aloisio}}, \bibinfo {author} {\bibfnamefont
  {E.}~\bibnamefont {Amato}}, \bibinfo {author} {\bibfnamefont
  {P.}~\bibnamefont {Attin{\'a}}}, \bibinfo {author} {\bibfnamefont
  {M.}~\bibnamefont {Axelsson}}, \bibinfo {author} {\bibfnamefont
  {L.}~\bibnamefont {Baldini}}, \bibinfo {author} {\bibfnamefont
  {S.}~\bibnamefont {Basso}}, \bibinfo {author} {\bibfnamefont
  {S.}~\bibnamefont {Bianchi}}, \bibinfo {author} {\bibfnamefont
  {P.}~\bibnamefont {Blasi}}, \bibinfo {author} {\bibfnamefont
  {J.}~\bibnamefont {Bregeon}}, \bibinfo {author} {\bibfnamefont
  {A.}~\bibnamefont {Brez}}, \bibinfo {author} {\bibfnamefont {N.}~\bibnamefont
  {Bucciantini}}, \bibinfo {author} {\bibfnamefont {L.}~\bibnamefont
  {Burderi}}, \bibinfo {author} {\bibfnamefont {V.}~\bibnamefont {Burwitz}},
  \bibinfo {author} {\bibfnamefont {P.}~\bibnamefont {Casella}}, \bibinfo
  {author} {\bibfnamefont {E.}~\bibnamefont {Churazov}}, \bibinfo {author}
  {\bibfnamefont {M.}~\bibnamefont {Civitani}}, \bibinfo {author}
  {\bibfnamefont {S.}~\bibnamefont {Covino}}, \bibinfo {author} {\bibfnamefont
  {R.~M.}\ \bibnamefont {Curado~da Silva}}, \bibinfo {author} {\bibfnamefont
  {G.}~\bibnamefont {Cusumano}}, \bibinfo {author} {\bibfnamefont
  {M.}~\bibnamefont {Dadina}}, \bibinfo {author} {\bibfnamefont
  {F.}~\bibnamefont {D'Amico}}, \bibinfo {author} {\bibfnamefont
  {A.}~\bibnamefont {De~Rosa}}, \bibinfo {author} {\bibfnamefont
  {S.}~\bibnamefont {Di~Cosimo}}, \bibinfo {author} {\bibfnamefont
  {G.}~\bibnamefont {Di~Persio}}, \bibinfo {author} {\bibfnamefont
  {T.}~\bibnamefont {Di~Salvo}}, \bibinfo {author} {\bibfnamefont
  {M.}~\bibnamefont {Dovciak}}, \bibinfo {author} {\bibfnamefont
  {R.}~\bibnamefont {Elsner}}, \bibinfo {author} {\bibfnamefont {C.~J.}\
  \bibnamefont {Eyles}}, \bibinfo {author} {\bibfnamefont {A.~C.}\ \bibnamefont
  {Fabian}}, \bibinfo {author} {\bibfnamefont {S.}~\bibnamefont {Fabiani}},
  \bibinfo {author} {\bibfnamefont {H.}~\bibnamefont {Feng}}, \bibinfo {author}
  {\bibfnamefont {S.}~\bibnamefont {Giarrusso}}, \bibinfo {author}
  {\bibfnamefont {R.~W.}\ \bibnamefont {Goosmann}}, \bibinfo {author}
  {\bibfnamefont {P.}~\bibnamefont {Grandi}}, \bibinfo {author} {\bibfnamefont
  {N.}~\bibnamefont {Grosso}}, \bibinfo {author} {\bibfnamefont
  {G.}~\bibnamefont {Israel}}, \bibinfo {author} {\bibfnamefont
  {M.}~\bibnamefont {Jackson}}, \bibinfo {author} {\bibfnamefont
  {P.}~\bibnamefont {Kaaret}}, \bibinfo {author} {\bibfnamefont
  {V.}~\bibnamefont {Karas}}, \bibinfo {author} {\bibfnamefont
  {M.}~\bibnamefont {Kuss}}, \bibinfo {author} {\bibfnamefont {D.}~\bibnamefont
  {Lai}}, \bibinfo {author} {\bibfnamefont {G.~L.}\ \bibnamefont {Rosa}},
  \bibinfo {author} {\bibfnamefont {J.}~\bibnamefont {Larsson}}, \bibinfo
  {author} {\bibfnamefont {S.}~\bibnamefont {Larsson}}, \bibinfo {author}
  {\bibfnamefont {L.}~\bibnamefont {Latronico}}, \bibinfo {author}
  {\bibfnamefont {A.}~\bibnamefont {Maggio}}, \bibinfo {author} {\bibfnamefont
  {J.}~\bibnamefont {Maia}}, \bibinfo {author} {\bibfnamefont {F.}~\bibnamefont
  {Marin}}, \bibinfo {author} {\bibfnamefont {M.~M.}\ \bibnamefont {Massai}},
  \bibinfo {author} {\bibfnamefont {T.}~\bibnamefont {Mineo}}, \bibinfo
  {author} {\bibfnamefont {M.}~\bibnamefont {Minuti}}, \bibinfo {author}
  {\bibfnamefont {E.}~\bibnamefont {Moretti}}, \bibinfo {author} {\bibfnamefont
  {F.}~\bibnamefont {Muleri}}, \bibinfo {author} {\bibfnamefont {S.~L.}\
  \bibnamefont {O'Dell}}, \bibinfo {author} {\bibfnamefont {G.}~\bibnamefont
  {Pareschi}}, \bibinfo {author} {\bibfnamefont {G.}~\bibnamefont {Peres}},
  \bibinfo {author} {\bibfnamefont {M.}~\bibnamefont {Pesce}}, \bibinfo
  {author} {\bibfnamefont {P.-O.}\ \bibnamefont {Petrucci}}, \bibinfo {author}
  {\bibfnamefont {M.}~\bibnamefont {Pinchera}}, \bibinfo {author}
  {\bibfnamefont {D.}~\bibnamefont {Porquet}}, \bibinfo {author} {\bibfnamefont
  {B.}~\bibnamefont {Ramsey}}, \bibinfo {author} {\bibfnamefont
  {N.}~\bibnamefont {Rea}}, \bibinfo {author} {\bibfnamefont {F.}~\bibnamefont
  {Reale}}, \bibinfo {author} {\bibfnamefont {J.~M.}\ \bibnamefont {Rodrigo}},
  \bibinfo {author} {\bibfnamefont {A.}~\bibnamefont
  {R{\'o}{\.{z}}a{\'{n}}ska}}, \bibinfo {author} {\bibfnamefont
  {A.}~\bibnamefont {Rubini}}, \bibinfo {author} {\bibfnamefont
  {P.}~\bibnamefont {Rudawy}}, \bibinfo {author} {\bibfnamefont
  {F.}~\bibnamefont {Ryde}}, \bibinfo {author} {\bibfnamefont {M.}~\bibnamefont
  {Salvati}}, \bibinfo {author} {\bibfnamefont {V.~A.}\ \bibnamefont
  {de~Santiago}}, \bibinfo {author} {\bibfnamefont {S.}~\bibnamefont
  {Sazonov}}, \bibinfo {author} {\bibfnamefont {C.}~\bibnamefont {Sgr{\'o}}},
  \bibinfo {author} {\bibfnamefont {E.}~\bibnamefont {Silver}}, \bibinfo
  {author} {\bibfnamefont {G.}~\bibnamefont {Spandre}}, \bibinfo {author}
  {\bibfnamefont {D.}~\bibnamefont {Spiga}}, \bibinfo {author} {\bibfnamefont
  {L.}~\bibnamefont {Stella}}, \bibinfo {author} {\bibfnamefont
  {T.}~\bibnamefont {Tamagawa}}, \bibinfo {author} {\bibfnamefont
  {F.}~\bibnamefont {Tamborra}}, \bibinfo {author} {\bibfnamefont
  {F.}~\bibnamefont {Tavecchio}}, \bibinfo {author} {\bibfnamefont
  {T.}~\bibnamefont {Teixeira~Dias}}, \bibinfo {author} {\bibfnamefont
  {M.}~\bibnamefont {van Adelsberg}}, \bibinfo {author} {\bibfnamefont
  {K.}~\bibnamefont {Wu}}, \ and\ \bibinfo {author} {\bibfnamefont
  {S.}~\bibnamefont {Zane}},\ }\href@noop {} {\bibfield  {journal} {\bibinfo
  {journal} {Experimental Astronomy}\ }\textbf {\bibinfo {volume} {36}},\
  \bibinfo {pages} {523} (\bibinfo {year} {2013})}\BibitemShut {NoStop}%
\bibitem [{201(2013)}]{2013SPIE.8777E..0WA}%
  \BibitemOpen
  \href@noop {} {\emph {\bibinfo {title} {{X-ray optic developments at NASA's
  MSFC}}}}\ (\bibinfo {organization} {The Univ. of Alabama in Huntsville
  (United States)},\ \bibinfo {year} {2013})\BibitemShut {NoStop}%
\bibitem [{\citenamefont {Wilson-Hodge}\ \emph {et~al.}(2015)\citenamefont
  {Wilson-Hodge}, \citenamefont {Cherry}, \citenamefont {Case}, \citenamefont
  {Baumgartner}, \citenamefont {Beklen}, \citenamefont {Bhat}, \citenamefont
  {Briggs}, \citenamefont {Buehler}, \citenamefont {Camero-Arranz},
  \citenamefont {Connaughton}, \citenamefont {Diehl}, \citenamefont {Finger},
  \citenamefont {Gehrels}, \citenamefont {Greiner}, \citenamefont {Harrison},
  \citenamefont {Hays}, \citenamefont {Jahoda}, \citenamefont {Jenke},
  \citenamefont {Kippen}, \citenamefont {Kouveliotou}, \citenamefont {Krimm},
  \citenamefont {Kuulkers}, \citenamefont {Madsen}, \citenamefont {Markwardt},
  \citenamefont {Meegan}, \citenamefont {Natalucci}, \citenamefont {Paciesas},
  \citenamefont {Preece}, \citenamefont {Rodi}, \citenamefont {Shaposhnikov},
  \citenamefont {Skinner}, \citenamefont {Swartz}, \citenamefont {von
  Kienlin},\ and\ \citenamefont {Zhang}}]{2015AAS...22521401W}%
  \BibitemOpen
  \bibfield  {author} {\bibinfo {author} {\bibfnamefont {C.}~\bibnamefont
  {Wilson-Hodge}}, \bibinfo {author} {\bibfnamefont {M.~L.}\ \bibnamefont
  {Cherry}}, \bibinfo {author} {\bibfnamefont {G.~L.}\ \bibnamefont {Case}},
  \bibinfo {author} {\bibfnamefont {W.~H.}\ \bibnamefont {Baumgartner}},
  \bibinfo {author} {\bibfnamefont {E.}~\bibnamefont {Beklen}}, \bibinfo
  {author} {\bibfnamefont {N.~P.}\ \bibnamefont {Bhat}}, \bibinfo {author}
  {\bibfnamefont {M.~S.}\ \bibnamefont {Briggs}}, \bibinfo {author}
  {\bibfnamefont {R.}~\bibnamefont {Buehler}}, \bibinfo {author} {\bibfnamefont
  {A.}~\bibnamefont {Camero-Arranz}}, \bibinfo {author} {\bibfnamefont
  {V.}~\bibnamefont {Connaughton}}, \bibinfo {author} {\bibfnamefont
  {R.}~\bibnamefont {Diehl}}, \bibinfo {author} {\bibfnamefont {M.~H.}\
  \bibnamefont {Finger}}, \bibinfo {author} {\bibfnamefont {N.}~\bibnamefont
  {Gehrels}}, \bibinfo {author} {\bibfnamefont {J.}~\bibnamefont {Greiner}},
  \bibinfo {author} {\bibfnamefont {F.}~\bibnamefont {Harrison}}, \bibinfo
  {author} {\bibfnamefont {E.~A.}\ \bibnamefont {Hays}}, \bibinfo {author}
  {\bibfnamefont {K.}~\bibnamefont {Jahoda}}, \bibinfo {author} {\bibfnamefont
  {P.}~\bibnamefont {Jenke}}, \bibinfo {author} {\bibfnamefont {R.~M.}\
  \bibnamefont {Kippen}}, \bibinfo {author} {\bibfnamefont {C.}~\bibnamefont
  {Kouveliotou}}, \bibinfo {author} {\bibfnamefont {H.~A.}\ \bibnamefont
  {Krimm}}, \bibinfo {author} {\bibfnamefont {E.}~\bibnamefont {Kuulkers}},
  \bibinfo {author} {\bibfnamefont {K.}~\bibnamefont {Madsen}}, \bibinfo
  {author} {\bibfnamefont {C.}~\bibnamefont {Markwardt}}, \bibinfo {author}
  {\bibfnamefont {C.~A.}\ \bibnamefont {Meegan}}, \bibinfo {author}
  {\bibfnamefont {L.}~\bibnamefont {Natalucci}}, \bibinfo {author}
  {\bibfnamefont {W.~S.}\ \bibnamefont {Paciesas}}, \bibinfo {author}
  {\bibfnamefont {R.~D.}\ \bibnamefont {Preece}}, \bibinfo {author}
  {\bibfnamefont {J.}~\bibnamefont {Rodi}}, \bibinfo {author} {\bibfnamefont
  {N.}~\bibnamefont {Shaposhnikov}}, \bibinfo {author} {\bibfnamefont {G.~K.}\
  \bibnamefont {Skinner}}, \bibinfo {author} {\bibfnamefont {D.~A.}\
  \bibnamefont {Swartz}}, \bibinfo {author} {\bibfnamefont {A.}~\bibnamefont
  {von Kienlin}}, \ and\ \bibinfo {author} {\bibfnamefont {X.-L.}\ \bibnamefont
  {Zhang}},\ }\href@noop {} {\bibfield  {journal} {\bibinfo  {journal}
  {American Astronomical Society}\ }\textbf {\bibinfo {volume} {225}},\
  \bibinfo {pages} {214.01} (\bibinfo {year} {2015})}\BibitemShut {NoStop}%
\bibitem [{\citenamefont {{Zarei, M}}\ \emph {et~al.}(2010)\citenamefont
  {{Zarei, M}}, \citenamefont {{Bavarsad, E}}, \citenamefont {{Haghighat, M}},
  \citenamefont {{Mohammadi, R}}, \citenamefont {{Motie, I}},\ and\
  \citenamefont {{Rezaei, Z}}}]{2010PhRvD..81h4035Z}%
  \BibitemOpen
  \bibfield  {author} {\bibinfo {author} {\bibnamefont {{Zarei, M}}}, \bibinfo
  {author} {\bibnamefont {{Bavarsad, E}}}, \bibinfo {author} {\bibnamefont
  {{Haghighat, M}}}, \bibinfo {author} {\bibnamefont {{Mohammadi, R}}},
  \bibinfo {author} {\bibnamefont {{Motie, I}}}, \ and\ \bibinfo {author}
  {\bibnamefont {{Rezaei, Z}}},\ }\href@noop {} {\bibfield  {journal} {\bibinfo
   {journal} {Physical Review D}\ }\textbf {\bibinfo {volume} {81}},\ \bibinfo
  {pages} {084035} (\bibinfo {year} {2010})}\BibitemShut {NoStop}%
\bibitem [{\citenamefont {{Tizchang}}\ \emph {et~al.}(2017)\citenamefont
  {{Tizchang}}, \citenamefont {{Batebi}}, \citenamefont {{Haghighat}},\ and\
  \citenamefont {{Mohammadi}}}]{2017JHEP...02..003T}%
  \BibitemOpen
  \bibfield  {author} {\bibinfo {author} {\bibfnamefont {S.}~\bibnamefont
  {{Tizchang}}}, \bibinfo {author} {\bibfnamefont {S.}~\bibnamefont
  {{Batebi}}}, \bibinfo {author} {\bibfnamefont {M.}~\bibnamefont
  {{Haghighat}}}, \ and\ \bibinfo {author} {\bibfnamefont {R.}~\bibnamefont
  {{Mohammadi}}},\ }\href {\doibase 10.1007/JHEP02(2017)003} {\bibfield
  {journal} {\bibinfo  {journal} {Journal of High Energy Physics}\ }\textbf
  {\bibinfo {volume} {2}},\ \bibinfo {eid} {3} (\bibinfo {year} {2017})},\
  \Eprint {http://arxiv.org/abs/1608.01231} {arXiv:1608.01231 [hep-ph]}
  \BibitemShut {NoStop}%
\bibitem [{\citenamefont {Mohammadi}\ and\ \citenamefont
  {Xue}(2014)}]{Mohammadi:2013ksa}%
  \BibitemOpen
  \bibfield  {author} {\bibinfo {author} {\bibfnamefont {R.}~\bibnamefont
  {Mohammadi}}\ and\ \bibinfo {author} {\bibfnamefont {S.-S.}\ \bibnamefont
  {Xue}},\ }\href {\doibase 10.1016/j.physletb.2014.02.031} {\bibfield
  {journal} {\bibinfo  {journal} {Phys. Lett.}\ }\textbf {\bibinfo {volume}
  {B731}},\ \bibinfo {pages} {272} (\bibinfo {year} {2014})},\ \Eprint
  {http://arxiv.org/abs/1312.3862} {arXiv:1312.3862 [hep-ph]} \BibitemShut
  {NoStop}%
\bibitem [{\citenamefont {Khodagholizadeh}\ \emph {et~al.}(2014)\citenamefont
  {Khodagholizadeh}, \citenamefont {Mohammadi},\ and\ \citenamefont
  {Xue}}]{Khodagholizadeh:2014nfa}%
  \BibitemOpen
  \bibfield  {author} {\bibinfo {author} {\bibfnamefont {J.}~\bibnamefont
  {Khodagholizadeh}}, \bibinfo {author} {\bibfnamefont {R.}~\bibnamefont
  {Mohammadi}}, \ and\ \bibinfo {author} {\bibfnamefont {S.-S.}\ \bibnamefont
  {Xue}},\ }\href {\doibase 10.1103/PhysRevD.90.091301} {\bibfield  {journal}
  {\bibinfo  {journal} {Phys. Rev.}\ }\textbf {\bibinfo {volume} {D90}},\
  \bibinfo {pages} {091301} (\bibinfo {year} {2014})},\ \Eprint
  {http://arxiv.org/abs/1406.6213} {arXiv:1406.6213 [astro-ph.CO]} \BibitemShut
  {NoStop}%
\bibitem [{\citenamefont {Mohammadi}\ \emph {et~al.}(2014)\citenamefont
  {Mohammadi}, \citenamefont {Motie},\ and\ \citenamefont
  {Xue}}]{2014PhRvA..89f2111M}%
  \BibitemOpen
  \bibfield  {author} {\bibinfo {author} {\bibfnamefont {R.}~\bibnamefont
  {Mohammadi}}, \bibinfo {author} {\bibfnamefont {I.}~\bibnamefont {Motie}}, \
  and\ \bibinfo {author} {\bibfnamefont {S.-S.}\ \bibnamefont {Xue}},\
  }\href@noop {} {\bibfield  {journal} {\bibinfo  {journal} {Physical Review
  A}\ }\textbf {\bibinfo {volume} {89}},\ \bibinfo {pages} {062111} (\bibinfo
  {year} {2014})}\BibitemShut {NoStop}%
\bibitem [{\citenamefont {Motie}\ and\ \citenamefont
  {Xue}(2012)}]{Motie:2011az}%
  \BibitemOpen
  \bibfield  {author} {\bibinfo {author} {\bibfnamefont {I.}~\bibnamefont
  {Motie}}\ and\ \bibinfo {author} {\bibfnamefont {S.-S.}\ \bibnamefont
  {Xue}},\ }\href {\doibase 10.1209/0295-5075/100/17006} {\bibfield  {journal}
  {\bibinfo  {journal} {Europhys. Lett.}\ }\textbf {\bibinfo {volume} {100}},\
  \bibinfo {pages} {17006} (\bibinfo {year} {2012})},\ \Eprint
  {http://arxiv.org/abs/1104.3555} {arXiv:1104.3555 [hep-ph]} \BibitemShut
  {NoStop}%
\bibitem [{\citenamefont {Shakeri}\ \emph {et~al.}(2017)\citenamefont
  {Shakeri}, \citenamefont {Kalantari},\ and\ \citenamefont
  {Xue}}]{2017PhRvA..95a2108S}%
  \BibitemOpen
  \bibfield  {author} {\bibinfo {author} {\bibfnamefont {S.}~\bibnamefont
  {Shakeri}}, \bibinfo {author} {\bibfnamefont {S.~Z.}\ \bibnamefont
  {Kalantari}}, \ and\ \bibinfo {author} {\bibfnamefont {S.-S.}\ \bibnamefont
  {Xue}},\ }\href@noop {} {\bibfield  {journal} {\bibinfo  {journal} {Physical
  Review A}\ }\textbf {\bibinfo {volume} {95}},\ \bibinfo {pages} {012108}
  (\bibinfo {year} {2017})}\BibitemShut {NoStop}%
\bibitem [{\citenamefont {Chandrasekhar}(1960)}]{Chandrasekhar:1960tz}%
  \BibitemOpen
  \bibfield  {author} {\bibinfo {author} {\bibfnamefont {S.}~\bibnamefont
  {Chandrasekhar}},\ }\href@noop {} {\emph {\bibinfo {title} {{Radiative heat
  transfer}}}}\ (\bibinfo  {publisher} {Dover Publications},\ \bibinfo {year}
  {1960})\BibitemShut {NoStop}%
\bibitem [{\citenamefont {Jackson}(1975)}]{Jackson:1975up}%
  \BibitemOpen
  \bibfield  {author} {\bibinfo {author} {\bibfnamefont {W.~D.}\ \bibnamefont
  {Jackson}},\ }\href@noop {} {\emph {\bibinfo {title} {{Classical
  Electrodynamics, Wiley {\&} Sons Inc}}}}\ (\bibinfo  {publisher} {Capitolo 2
  Par. 9. Capitolo 2 Par},\ \bibinfo {year} {1975})\BibitemShut {NoStop}%
\bibitem [{\citenamefont {Rybicki}\ and\ \citenamefont
  {Lightman}(2008)}]{Rybicki:2008vo}%
  \BibitemOpen
  \bibfield  {author} {\bibinfo {author} {\bibfnamefont {G.~B.}\ \bibnamefont
  {Rybicki}}\ and\ \bibinfo {author} {\bibfnamefont {A.~P.}\ \bibnamefont
  {Lightman}},\ }\href@noop {} {\emph {\bibinfo {title} {{Radiative Processes
  in Astrophysics}}}}\ (\bibinfo  {publisher} {John Wiley {\&} Sons},\ \bibinfo
  {year} {2008})\BibitemShut {NoStop}%
\bibitem [{\citenamefont {Kosowsky}(1996)}]{1996AnPhy.246...49K}%
  \BibitemOpen
  \bibfield  {author} {\bibinfo {author} {\bibfnamefont {A.}~\bibnamefont
  {Kosowsky}},\ }\href@noop {} {\bibfield  {journal} {\bibinfo  {journal} {Ann.
  Phys. (USA)}\ }\textbf {\bibinfo {volume} {246}},\ \bibinfo {pages} {49}
  (\bibinfo {year} {1996})}\BibitemShut {NoStop}%
\bibitem [{\citenamefont {Alexander}\ \emph {et~al.}(2009)\citenamefont
  {Alexander}, \citenamefont {Ochoa},\ and\ \citenamefont
  {Kosowsky}}]{2009PhRvD..79f3524A}%
  \BibitemOpen
  \bibfield  {author} {\bibinfo {author} {\bibfnamefont {S.}~\bibnamefont
  {Alexander}}, \bibinfo {author} {\bibfnamefont {J.}~\bibnamefont {Ochoa}}, \
  and\ \bibinfo {author} {\bibfnamefont {A.}~\bibnamefont {Kosowsky}},\
  }\href@noop {} {\bibfield  {journal} {\bibinfo  {journal} {Physical Review
  D}\ }\textbf {\bibinfo {volume} {79}},\ \bibinfo {pages} {063524} (\bibinfo
  {year} {2009})}\BibitemShut {NoStop}%
\bibitem [{\citenamefont {{Euler, Hans}}(1936)}]{1936AnP...418..398E}%
  \BibitemOpen
  \bibfield  {author} {\bibinfo {author} {\bibnamefont {{Euler, Hans}}},\
  }\href@noop {} {\bibfield  {journal} {\bibinfo  {journal} {Annalen der
  Physik}\ }\textbf {\bibinfo {volume} {418}},\ \bibinfo {pages} {398}
  (\bibinfo {year} {1936})}\BibitemShut {NoStop}%
\bibitem [{\citenamefont {{Dicus, Duane A}}\ \emph {et~al.}(1998)\citenamefont
  {{Dicus, Duane A}}, \citenamefont {{Kao, Chung}},\ and\ \citenamefont
  {{Repko, Wayne W}}}]{1998PhRvD..57.2443D}%
  \BibitemOpen
  \bibfield  {author} {\bibinfo {author} {\bibnamefont {{Dicus, Duane A}}},
  \bibinfo {author} {\bibnamefont {{Kao, Chung}}}, \ and\ \bibinfo {author}
  {\bibnamefont {{Repko, Wayne W}}},\ }\href@noop {} {\bibfield  {journal}
  {\bibinfo  {journal} {Physical Review D (Particles}\ }\textbf {\bibinfo
  {volume} {57}},\ \bibinfo {pages} {2443} (\bibinfo {year}
  {1998})}\BibitemShut {NoStop}%
\bibitem [{\citenamefont {{Dunne}}(2012)}]{Dunne:2012hp}%
  \BibitemOpen
  \bibfield  {author} {\bibinfo {author} {\bibfnamefont {G.~V.}\ \bibnamefont
  {{Dunne}}},\ }\href {\doibase 10.1142/S0217751X12600044} {\bibfield
  {journal} {\bibinfo  {journal} {International Journal of Modern Physics A}\
  }\textbf {\bibinfo {volume} {27}},\ \bibinfo {eid} {1260004} (\bibinfo {year}
  {2012})},\ \Eprint {http://arxiv.org/abs/1202.1557} {arXiv:1202.1557
  [hep-th]} \BibitemShut {NoStop}%
\bibitem [{\citenamefont {Beloborodov}(2002)}]{2002ApJ...566L..85B}%
  \BibitemOpen
  \bibfield  {author} {\bibinfo {author} {\bibfnamefont {A.~M.}\ \bibnamefont
  {Beloborodov}},\ }\href@noop {} {\bibfield  {journal} {\bibinfo  {journal}
  {The Astrophysical Journal}\ }\textbf {\bibinfo {volume} {566}},\ \bibinfo
  {pages} {L85} (\bibinfo {year} {2002})}\BibitemShut {NoStop}%
\bibitem [{\citenamefont {{Marshall}}\ and\ \citenamefont
  {{Schulz}}(2014)}]{2014mbhe.conf..288M}%
  \BibitemOpen
  \bibfield  {author} {\bibinfo {author} {\bibfnamefont {H.~L.}\ \bibnamefont
  {{Marshall}}}\ and\ \bibinfo {author} {\bibfnamefont {N.~S.}\ \bibnamefont
  {{Schulz}}},\ }in\ \href {\doibase 10.14311/APP.2014.01.0288} {\emph
  {\bibinfo {booktitle} {Multifrequency Behaviour of High Energy Cosmic
  Sources}}}\ (\bibinfo {year} {2014})\ pp.\ \bibinfo {pages}
  {288--292}\BibitemShut {NoStop}%
\bibitem [{\citenamefont {Kubo}\ and\ \citenamefont
  {Nagata}(1983)}]{Kubo:1983fk}%
  \BibitemOpen
  \bibfield  {author} {\bibinfo {author} {\bibfnamefont {H.}~\bibnamefont
  {Kubo}}\ and\ \bibinfo {author} {\bibfnamefont {R.}~\bibnamefont {Nagata}},\
  }\href@noop {} {\bibfield  {journal} {\bibinfo  {journal} {JOSA}\ }\textbf
  {\bibinfo {volume} {73}},\ \bibinfo {pages} {1719} (\bibinfo {year}
  {1983})}\BibitemShut {NoStop}%
\bibitem [{\citenamefont {{Taverna, R}}\ \emph {et~al.}(2015)\citenamefont
  {{Taverna, R}}, \citenamefont {{Turolla, R}}, \citenamefont {{Gonzalez
  Caniulef, D}}, \citenamefont {{Zane, S}}, \citenamefont {{Muleri, F}},\ and\
  \citenamefont {{Soffitta, P}}}]{Taverna:2015uu}%
  \BibitemOpen
  \bibfield  {author} {\bibinfo {author} {\bibnamefont {{Taverna, R}}},
  \bibinfo {author} {\bibnamefont {{Turolla, R}}}, \bibinfo {author}
  {\bibnamefont {{Gonzalez Caniulef, D}}}, \bibinfo {author} {\bibnamefont
  {{Zane, S}}}, \bibinfo {author} {\bibnamefont {{Muleri, F}}}, \ and\ \bibinfo
  {author} {\bibnamefont {{Soffitta, P}}},\ }\href@noop {} {\bibfield
  {journal} {\bibinfo  {journal} {Monthly Notices of the Royal Astronomical
  Society}\ }\textbf {\bibinfo {volume} {454}},\ \bibinfo {pages} {3254}
  (\bibinfo {year} {2015})}\BibitemShut {NoStop}%
\bibitem [{\citenamefont {Kislat}\ \emph {et~al.}(2015)\citenamefont {Kislat},
  \citenamefont {Clark}, \citenamefont {Beilicke},\ and\ \citenamefont
  {Krawczynski}}]{Kislat:2015vt}%
  \BibitemOpen
  \bibfield  {author} {\bibinfo {author} {\bibfnamefont {F.}~\bibnamefont
  {Kislat}}, \bibinfo {author} {\bibfnamefont {B.}~\bibnamefont {Clark}},
  \bibinfo {author} {\bibfnamefont {M.}~\bibnamefont {Beilicke}}, \ and\
  \bibinfo {author} {\bibfnamefont {H.}~\bibnamefont {Krawczynski}},\
  }\href@noop {} {\bibfield  {journal} {\bibinfo  {journal} {Astroparticle
  Physics}\ } (\bibinfo {year} {2015})}\BibitemShut {NoStop}%
\bibitem [{\citenamefont {van Adelsberg}\ and\ \citenamefont
  {Perna}(2009)}]{2009MNRAS.399.1523V}%
  \BibitemOpen
  \bibfield  {author} {\bibinfo {author} {\bibfnamefont {M.}~\bibnamefont {van
  Adelsberg}}\ and\ \bibinfo {author} {\bibfnamefont {R.}~\bibnamefont
  {Perna}},\ }\href@noop {} {\bibfield  {journal} {\bibinfo  {journal} {Monthly
  Notices of the Royal Astronomical Society}\ }\textbf {\bibinfo {volume}
  {399}},\ \bibinfo {pages} {1523} (\bibinfo {year} {2009})}\BibitemShut
  {NoStop}%
\bibitem [{\citenamefont {Taverna}\ \emph {et~al.}(2014)\citenamefont
  {Taverna}, \citenamefont {Muleri}, \citenamefont {Turolla}, \citenamefont
  {Soffitta}, \citenamefont {Fabiani},\ and\ \citenamefont
  {Nobili}}]{2014MNRAS.438.1686T}%
  \BibitemOpen
  \bibfield  {author} {\bibinfo {author} {\bibfnamefont {R.}~\bibnamefont
  {Taverna}}, \bibinfo {author} {\bibfnamefont {F.}~\bibnamefont {Muleri}},
  \bibinfo {author} {\bibfnamefont {R.}~\bibnamefont {Turolla}}, \bibinfo
  {author} {\bibfnamefont {P.}~\bibnamefont {Soffitta}}, \bibinfo {author}
  {\bibfnamefont {S.}~\bibnamefont {Fabiani}}, \ and\ \bibinfo {author}
  {\bibfnamefont {L.}~\bibnamefont {Nobili}},\ }\href@noop {} {\bibfield
  {journal} {\bibinfo  {journal} {Monthly Notices of the Royal Astronomical
  Society}\ }\textbf {\bibinfo {volume} {438}},\ \bibinfo {pages} {1686}
  (\bibinfo {year} {2014})}\BibitemShut {NoStop}%
\bibitem [{\citenamefont {Pechenick}\ \emph {et~al.}(1983)\citenamefont
  {Pechenick}, \citenamefont {Ftaclas},\ and\ \citenamefont
  {Cohen}}]{1983ApJ...274..846P}%
  \BibitemOpen
  \bibfield  {author} {\bibinfo {author} {\bibfnamefont {K.~R.}\ \bibnamefont
  {Pechenick}}, \bibinfo {author} {\bibfnamefont {C.}~\bibnamefont {Ftaclas}},
  \ and\ \bibinfo {author} {\bibfnamefont {J.~M.}\ \bibnamefont {Cohen}},\
  }\href@noop {} {\bibfield  {journal} {\bibinfo  {journal} {Astrophysical
  Journal}\ }\textbf {\bibinfo {volume} {274}},\ \bibinfo {pages} {846}
  (\bibinfo {year} {1983})}\BibitemShut {NoStop}%
\bibitem [{\citenamefont {Turolla}\ and\ \citenamefont
  {Nobili}(2013)}]{2013ApJ...768..147T}%
  \BibitemOpen
  \bibfield  {author} {\bibinfo {author} {\bibfnamefont {R.}~\bibnamefont
  {Turolla}}\ and\ \bibinfo {author} {\bibfnamefont {L.}~\bibnamefont
  {Nobili}},\ }\href@noop {} {\bibfield  {journal} {\bibinfo  {journal} {The
  Astrophysical Journal}\ }\textbf {\bibinfo {volume} {768}},\ \bibinfo {pages}
  {147} (\bibinfo {year} {2013})}\BibitemShut {NoStop}%
\bibitem [{\citenamefont {Wiktorowicz}\ \emph {et~al.}(2015)\citenamefont
  {Wiktorowicz}, \citenamefont {Ramirez-Ruiz}, \citenamefont {Illing},\ and\
  \citenamefont {Nofi}}]{2015AAS...22542101W}%
  \BibitemOpen
  \bibfield  {author} {\bibinfo {author} {\bibfnamefont {S.}~\bibnamefont
  {Wiktorowicz}}, \bibinfo {author} {\bibfnamefont {E.}~\bibnamefont
  {Ramirez-Ruiz}}, \bibinfo {author} {\bibfnamefont {R.~M.~E.}\ \bibnamefont
  {Illing}}, \ and\ \bibinfo {author} {\bibfnamefont {L.}~\bibnamefont
  {Nofi}},\ }\href@noop {} {\bibfield  {journal} {\bibinfo  {journal} {American
  Astronomical Society}\ }\textbf {\bibinfo {volume} {225}},\ \bibinfo {pages}
  {421.01} (\bibinfo {year} {2015})}\BibitemShut {NoStop}%
\bibitem [{\citenamefont {Fabiani}\ and\ \citenamefont
  {Muleri}(2014)}]{2014axp..book.....F}%
  \BibitemOpen
  \bibfield  {author} {\bibinfo {author} {\bibfnamefont {S.}~\bibnamefont
  {Fabiani}}\ and\ \bibinfo {author} {\bibfnamefont {F.}~\bibnamefont
  {Muleri}},\ }\href@noop {} {\bibfield  {journal} {\bibinfo  {journal}
  {Astronomical X-Ray Polarimetry by Sergio Fabiani and Fabio Muleri Foreword
  by Paolo Soffitta Introduction by Ronaldo Bellanzini}\ } (\bibinfo {year}
  {2014})}\BibitemShut {NoStop}%
\end{thebibliography}
\end{document}